\documentclass[aps,prd,showpacs,eqsecnum,twocolumn,superscriptaddress,a4paper,nofootinbib]
{revtex4-1}
\usepackage{amsmath,amssymb,graphicx,color}
\usepackage{bm}
\usepackage{epsf}

\begin{document}

\title{Alternative possibility of GW190521: Gravitational waves from
  high-mass black hole-disk systems}

\author{Masaru Shibata} 
\affiliation{Max Planck Institute for
  Gravitational Physics (Albert Einstein Institute), Am M{\"u}hlenberg 1,
  Potsdam-Golm 14476, Germany}
\affiliation{Center for Gravitational Physics, Yukawa Institute for Theoretical
  Physics, Kyoto University, Kyoto, 606-8502, Japan}

\author{Kenta Kiuchi}
\affiliation{Max Planck Institute for
  Gravitational Physics (Albert Einstein Institute), Am M{\"u}hlenberg 1,
  Potsdam-Golm 14476, Germany}
\affiliation{Center for Gravitational Physics, Yukawa Institute for Theoretical
  Physics, Kyoto University, Kyoto, 606-8502, Japan}

\author{Sho Fujibayashi} 
\affiliation{Max Planck Institute for
  Gravitational Physics (Albert Einstein Institute), Am M{\"u}hlenberg 1,
  Potsdam-Golm 14476, Germany}

\author{Yuichiro Sekiguchi} \affiliation{Department of Physics, Toho
  University, Funabashi, Chiba 274-8510, Japan}

\date{\today}
\newcommand{\beq}{\begin{equation}}
\newcommand{\eeq}{\end{equation}}
\newcommand{\beqn}{\begin{eqnarray}}
\newcommand{\eeqn}{\end{eqnarray}}
\newcommand{\pa}{\partial}
\newcommand{\vp}{\varphi}
\newcommand{\varep}{\varepsilon}
\newcommand{\ep}{\epsilon}
\newcommand{\comp}{(M/R)_\infty}
\begin{abstract}

We evolve high-mass disks of mass $15$--$50M_\odot$ orbiting a
$50M_\odot$ spinning black hole in the framework of numerical
relativity. Such high-mass systems could be an outcome during the
collapse of rapidly-rotating very-massive stars. The massive disks are
dynamically unstable to the so-called one-armed spiral-shape
deformation with the maximum fractional density-perturbation of
$\delta \rho/\rho \agt 0.1$, and hence, high-amplitude gravitational
waves are emitted.  The waveforms are characterized by an initial
high-amplitude burst with the frequency of $\sim 40$--$50$\,Hz and the
maximum amplitude of $(1$--$10)\times 10^{-22}$ at the hypothetical
distance of 100\,Mpc and by a subsequent low-amplitude quasi-periodic
oscillation. We illustrate that the waveforms in our models with a
wide range of the disk mass resemble that of GW190521. We also point
out that gravitational waves from rapidly-rotating very-massive stars
can be the source for 3rd-generation gravitational-wave detectors for
exploring the formation process of rapidly-spinning high-mass black
holes of mass $\sim 50$--$100M_\odot$ in an early universe.

\end{abstract}


\maketitle

\section{Introduction}\label{sec1}

In their third observational run, Advanced LIGO and Advanced Virgo
announced the discovery of a high-mass binary black hole,
GW190521~\cite{190521}. The estimated mass of two black holes are
$m_1=85^{+21}_{-14}M_\odot$ and $m_2=66^{+17}_{-18}M_\odot$ with the
estimated distance to the source of $5.3^{+2.4}_{-2.6}$\,Gpc.  Stellar
evolution calculations predict that black holes of mass gap $\sim
50$--$120M_\odot$ should not be formed from the massive stars because
of the presence of pulsational pair instability and pair-instability
explosion for the massive-star evolution with a range of the initial
stellar mass~\cite{woosley17,woosley19,Yoshida16}. For this reason,
the announcement of GW190521 by the LIGO-Virgo collaboration has been
attracting much attention and several possibilities for forming the
black holes in the mass gap have been suggested~(see, e.g.,
Refs.~\cite{190521a,190521b,190521c,190521d,190521e}).

Subsequently Nitz and Capano~\cite{Nitz} reanalyzed the data with
different priors and waveform models from those in Ref.~\cite{190521}.
They found that more probable component mass might be
$m_1=168^{+15}_{-61}M_\odot$ and $m_2=16^{+33}_{-3}M_\odot$, and thus,
the mass of them might not be in the mass gap range of
$50$--$120M_\odot$.  A very-massive black hole of mass $\agt
150M_\odot$ could be naturally interpreted as a remnant of the
collapse of a very-massive star (see below). Another interesting point
in their work is that their maximum-likelihood waveform indicates that
after the burst waveform of a few cycles with the highest amplitude,
gravitational waves might be emitted: They show that with their
maximum-likelihood waveform in which the post-burst waves are present,
the signal-to-noise ratio is higher than in the absence of the
post-burst waves. For binary black hole mergers, the gravitational
waveforms after a peak amplitude is reached should be characterized by
a ringdown waveform and the amplitude should approach zero
quickly. The possible presence of gravitational waves after the
burst-waves emission suggests that the source of GW190521 might not be
the merger of high-mass binary black holes.

Besides the frequency of gravitational waves, the waveforms such as that
of GW190521, which are composed initially of burst waves of a few wave
cycles and subsequently of a small-amplitude oscillation, have been
often reported in the results of numerical simulations for the systems
dynamically unstable to non-axisymmetric deformation. For example,
during the collapse of very rapidly rotating stars, the
non-axisymmetric deformation can occur and the resulting gravitational
waveform can be qualitatively similar to that of GW190521 (besides the
frequency: see, e.g., Fig.~15(b) of Ref.~\cite{SS05}). Thus, it would
not be non-sense to explore other astrophysical possibilities for
interpreting the observational results of GW190521, supposing that
post-burst gravitational waves indicated in Ref.~\cite{Nitz} may
indeed be real signals.

Based on this motivation, we perform a numerical-relativity simulation
for the system of a massive black hole and a massive disk.  We choose
the black hole of initial mass $M_{\rm BH,0}=50M_\odot$ and the disk
of rest mass $M_{\rm disk}=15$--$50M_\odot$ in this paper.  Here the
black-hole mass of $50M_\odot$ is chosen in order to reproduce the
typical frequency of gravitational waves for GW190521 which is $\sim
50$\,Hz~\cite{190521} (note that the frequency is approximately
proportional to $M_{\rm BH,0}^{-1}$).  Such a high-mass black hole-disk
system could be formed during the collapse of very-massive metal-poor
stars of initial mass larger than $\sim
190$--$260M_\odot$~\cite{UN02,HW02,Yoon2012,Yoon2015,Takahashi16,Takahashi18,FWH01}.
Here the critical initial mass depends on the angular momentum of the
progenitor stars~\cite{Yoon2012,Takahashi18}. For these stars, the
mass of the carbon-oxygen core formed at the final stage of the
stellar evolution is $M_{\rm CO}\agt 125M_\odot$ and the collapse is
triggered by the pair instability. The collapse of such heavy cores
leads to the direct formation of a high-mass black hole, avoiding the
pair-instability explosion, while for $70M_\odot \alt M_{\rm CO} \alt
125M_\odot$, a pair-instability supernova is the final outcome,
leaving no compact objects at the center~\cite{Takahashi18,Yoon2012}.

Here, it is important to emphasize that the mass of the black hole
during its formation may not be always equal to the entire mass of the
carbon-oxygen core and it could be smaller than $M_{\rm CO}$ in the
presence of rapid rotation. One reason for this is that a fraction of
the mass can form disks (and ejecta from the disks). Indeed, our
previous work in numerical relativity shows that this is the case:
Reference~\cite{uchida1} illustrates that after the collapse of a
moderately rapidly-rotating very-massive star of the initial mass
$320M_\odot$ and $M_{\rm CO}\approx 150M_\odot$, which is obtained by
a stellar evolution calculation~\cite{Takahashi18}, the outcome is a
black hole of mass $\sim 130M_\odot$ surrounded by a disk of mass
$\sim 20M_\odot$. Our subsequent work further illustrates that in the
presence of the neutrino cooling effect which was absent in
Ref.~\cite{uchida1}, the initial mass of the black hole formed during
the early stage of the collapse is likely to be substantially smaller
than $M_{\rm CO}$~\cite{uchida2} because the collapse of the central
region is significantly accelerated in a runaway manner toward the
earlier formation of a black hole.  The temporal mass of the black
hole and disk formed during the collapse could also depend strongly on
the angular velocity profile of the pre-collapse carbon-oxygen core,
which is not well understood by the stellar evolution calculations.
Thus, in this paper, we suppose the case in which a high angular
momentum is present in the carbon-oxygen core just prior to the
collapse, and a system of a black hole and a disk, for both of which
the mass is several tens of solar mass, could be formed during the
collapse of rapidly-rotating very-massive stellar cores.

By numerically evolving high-mass black hole-disk systems as a
plausible outcome formed during the collapse of rapidly-rotating
very-massive stars, we show that the massive disks in such systems are
dynamically unstable to the one-armed spiral-shape deformation as
often found in previous works for low-mass
disks~\cite{Hawley91,Zink07,Oleg11,Kiuchi11,Wessel} (but see also
Ref.~\cite{Bugli} for the possible importance of the magnetic field
effects for low-mass disks). Then, we demonstrate that gravitational
waves emitted from the non-linearly perturbed disks are indeed
composed of a high-amplitude initial burst and a subsequent
low-amplitude quasi-periodic oscillation.  For $M_{\rm
  BH,0}=50M_\odot$, the frequency of the initial burst gravitational
waves is $\sim 40$--$50$\,Hz for a wide range of the disk mass (for
higher disk mass, the frequency is lower), and we find that the
waveforms for several models are similar to the waveforms shown in
Refs.~\cite{190521,Nitz}, depending only weakly on the disk mass.

We note that during an early formation stage of a black hole and a
massive disk in the stellar collapse, the configuration of the system
is likely to be approximately axisymmetric until the onset of the
one-armed spiral-shape deformation. The amplitude of gravitational
waves emitted in the (approximately) axisymmetric stage could not be
as high as that emitted from the non-axisymmetric system even at the
moment of the black-hole formation (see, e.g., Ref.~\cite{uchida2}).
The non-axisymmetric instability in the disk is likely to set in when
the growth timescale of the dynamical instability becomes as short as
the formation timescale of the disk, which would be approximately
equal to the local free-fall timescale.  The amplitude of
gravitational waves emitted by the non-axisymmetric deformation of the
massive disk should be by $\sim 1$ orders of magnitude higher than
that emitted during the axisymmetric stage (e.g.,
Refs.~\cite{SS05,Ott11}).  Thus, for low signal-to-noise ratio events,
only gravitational waves emitted in the non-axisymmetric stage would
be observable.  This motivates us to perform simulations focusing only
on the black hole-disk system for exploring gravitational waves from
the collapse of rapidly-rotating very-massive stars as a first step.

This paper is organized as follows. In Sec.~\ref{sec2}, we summarize
the set-up of the present numerical-relativity simulations.  In
Sec.~\ref{sec3}, the unstable evolution of the high-mass disk
surrounding the high-mass black hole and resulting gravitational
waveforms are presented.  Section~\ref{sec4} is devoted to
discussions. Throughout this paper, $G$ and $c$ denote the
gravitational constant and the speed of light, respectively.

\section{Set-up of numerical simulations}\label{sec2}

The set-up of the present numerical simulations is essentially the
same as that in our previous papers on axisymmetric simulations for
the massive black hole-disk systems~\cite{Fujiba20,Fujiba21}, but in
this paper, we perform three-dimensional hydrodynamics simulations.
We numerically solve Einstein's equation and the hydrodynamics
equations.  Einstein's equation is solved using the original version
of the Baumgarte-Shapiro-Shibata-Nakamura formalism~\cite{BSSN}
together with the puncture formulation~\cite{puncture}, Z4c constraint
propagation prescription~\cite{Z4c}, and 6th-order Kreiss-Oliger
dissipation.  For hydrodynamics, we do not consider the neutrino
emission and weak interaction throughout this paper, although our
implementation can take into account the neutrino effects in the same
manner as in Refs.~\cite{Fujiba2018,Fujiba2020}.  In the presence of
the neutrino emission, the disk shrinks by the neutrino cooling and
its compactness is increased.  This effect makes the dependence of the
growth timescale of the disk instability and resulting gravitational
waveforms on the disk morphology obscure. Thus we decided to switch
off the neutrino effects in this work. In this prescription, only the
advection part for the equations of the electron fraction, $Y_e$, is taken
into account.

The quantities of black holes (mass and spin) are determined from
their area and circumferential radii of apparent horizons~\cite{K10},
assuming that these quantities are written as functions of the mass
and spin in the same formulation as in the vacuum Kerr black hole.
Gravitational waves are derived through the extraction of the outgoing
component of the complex Weyl scalar~\cite{SACRA}.  The wave
extraction is performed for coordinate spheres of radius
19000--35000\,km and we checked that the waveform depends very weakly
on the extraction radius. In Sec.~\ref{sec3}, we present the waveforms
extracted at 33000\,km($\approx 450M_{\rm BH,0}$). To obtain
gravitational waves from the complex Weyl scalar, we employ the
Reisswig-Pollney prescription~\cite{RP11} with the cutoff frequency of
8\,Hz for the $m=2$ mode.

As in our previous work~\cite{Fujiba20,Fujiba21}, we employ a
tabulated equation of state based on the DD2 equation of
state~\cite{DD2} for a relatively high-density part and the Timmes
(Helmholtz) equation of state for the low-density part~\cite{Timmes}.
In this tabulated equation of state, thermodynamics quantities such as
$\varep$, $P$, and $h$ are written as functions of $\rho$, $Y_e$, and
$T$ where $\varep$, $P$, $h(=c^2+\varep+P/\rho)$, $\rho$, and
$T$ are the specific internal energy, pressure, specific enthalpy,
rest-mass density, and matter temperature,
respectively.  We choose the lowest rest-mass density to be $0.1\,{\rm
  g/cm^3}$ in the table.



\begin{table}[t]
\caption{Initial conditions for the numerical simulation. Described
  are the model name, rest
  mass of the disk ($M_{\rm disk}$), the mass ratio of
  the disk to the black hole ($R_{\rm mass}:=M_{\rm disk}/M_{\rm BH,0}$), 
  dimensionless spin of the black
  hole ($\chi$), the coordinate radii at the inner and outer edges of
  the disk ($r_{\rm in}$ and $r_{\rm out}$), entropy per baryon ($s/k$
  with $k$ the Boltzmann's constant), the maximum density of the disk
  ($\rho_{\rm max}$), and the orbital period at the maximum density
  ($P_{\rm orb}$).  The units are $M_\odot$ for the mass, $GM_{\rm
    BH,0}/c^2 \approx 73.84(M_{\rm BH,0}/50M_\odot)$\,km for $r_{\rm
    in}$ and $r_{\rm out}$, $10^{10}\,{\rm g/cm^3}$ for $\rho_{\rm
    max}$, and $GM_{\rm BH,0}/c^3$ for $P_{\rm orb}$, respectively.  }
\begin{tabular}{ccccccccc} \hline
Model & $M_{\rm disk}$ & $R_{\rm mass}$ & ~~$\chi$~~
& ~$r_{\rm in}$~ & ~$r_{\rm out}$~ & ~$s/k$~ &~$\rho_{\rm max}$~&~$P_{\rm orb}$~
\\
 \hline \hline
L11 & 15.1 & 0.30 & 0.78 & 2.0 & 21.8 & 11 & 4.35 & 119 \\ 
L12 & 15.1 & 0.30 & 0.79 & 2.0 & 27.9 & 12 & 2.85 & 126 \\ 
J12 & 20.0 & 0.40 & 0.78 & 2.0 & 25.4 & 12 & 3.87 & 129 \\ 
J13 & 19.9 & 0.40 & 0.78 & 2.0 & 32.8 & 13 & 2.54 & 135 \\ 
K13 & 25.0 & 0.50 & 0.78 & 2.0 & 29.7 & 13 & 3.32 & 137 \\ 
K14 & 25.0 & 0.50 & 0.78 & 2.0 & 38.6 & 14 & 2.21 & 144 \\ 
M12 & 30.1 & 0.60 & 0.77 & 2.0 & 21.1 & 12 & 6.62 & 130 \\ 
M13 & 30.0 & 0.60 & 0.77 & 2.0 & 27.0 & 13 & 4.26 & 139 \\ 
M14 & 29.9 & 0.60 & 0.78 & 2.0 & 35.0 & 14 & 2.79 & 147 \\ 
M15 & 30.1 & 0.60 & 0.78 & 2.0 & 45.6 & 15 & 1.87 & 153 \\ 
H14 & 40.0 & 0.80 & 0.76 & 2.0 & 28.9 & 14 & 4.39 & 150 \\ 
H15 & 40.0 & 0.80 & 0.76 & 2.0 & 37.4 & 15 & 2.89 & 157 \\ 
H16 & 40.0 & 0.80 & 0.77 & 2.0 & 49.0 & 16 & 1.90 & 163 \\ 
E15 & 50.1 & 1.00 & 0.76 & 2.0 & 31.0 & 15 & 4.35 & 158 \\ 
E16 & 50.2 & 1.00 & 0.76 & 2.0 & 40.0 & 16 & 2.86 & 168 \\ 
E17 & 50.0 & 1.00 & 0.77 & 2.0 & 52.6 & 17 & 1.86 & 175 \\
\hline
\end{tabular}
\label{table1}
\end{table}

Axisymmetric equilibrium states for black hole-disk systems are
prepared as the initial conditions~\cite{Fujiba20,Fujiba21} using the method of
Ref.~\cite{Shibata2007}. As in Refs.~\cite{Fujiba20,Fujiba21}, we determine the
angular velocity profile from the relation
\beq
j \propto \Omega^{-n}, \label{eqj}
\eeq
where $j=c^{-2}h u_\varphi$ is the specific angular momentum.
$\Omega$ is the angular velocity defined by $u^\varphi/u^t$ with
$u^\mu$ the fluid four-velocity.  $n$ is a constant that determines
the profile of the angular velocity, for which we choose $n=1/7$
following our previous papers~\cite{Fujiba20,Fujiba21} to obtain a
profile of $\Omega$ close to the Keplerian one. Irrespective of the
value of $n$, the massive disk is geometrically mildly thick (see
appendix A). However the geometrical thickness is smaller for the
larger values of $n$, and thus, the disks employed in this paper are
geometrically thinner than those for the case of $j=$const $(n=0)$,
which are often used in the simulations for the black-hole accretion
disks.


For obtaining the initial conditions, we assume a plausible relation
between $\rho$ and $Y_e$ for the high-density matter in the same form
of $\rho(Y_e)$ as in Ref.~\cite{Fujiba20}.  For this model, the value
of $Y_e$ is 0.07 in a high density region with $\rho \geq
10^{11}\,{\rm g/cm^3}$, $0.07-(43/400)\log_{10}[\rho/(10^{11}{\rm
    g\,cm^{-3}})]$ for $10^7\,{\rm g/cm^3} \leq \rho \leq
10^{11}\,{\rm g/cm^3}$, and settles to $1/2$ for $\rho \leq
10^{7}\,{\rm g/cm^3}$, for which the effect of the electron degeneracy
to the neutron richness is weak.  In addition, we assume that the
specific entropy, $s$, is initially constant, in order to obtain the
first integral of the Euler equation easily.  For the high-mass disks,
we choose it $s/k=11$--17 in this paper where $k$ is the Boltzmann's
constant: By adjusting this value, the compactness for given disk mass
is controlled (for the higher entropy, the pressure for given
rest-mass density becomes higher, and hence, the disk becomes less
compact for a given value of the disk mass). An equilibrium state is
numerically obtained fixing the values of $r_{\rm in}/M_{\rm BH,0}$,
$r_{\rm out}/M_{\rm BH,0}$, $M_{\rm BH,0}$, and $s/k$ (see
Ref.~\cite{Fujiba20} for details).  To obtain a particular value for
the mass of the disk, we iteratively calculate equilibrium states
changing the value of $r_{\rm out}/M_{\rm BH,0}$ until the disk mass
relaxes to the desired value, while fixing $s/k$, $r_{\rm in}/M_{\rm
  BH,0}$, and $M_{\rm BH,0}$.

We note that our setting for the disk profile is idealized. For more
realistic work, it is obviously necessary to perform a simulation
started from rapidly-rotating very-massive progenitor stellar
cores. However, the collapse simulation with a wide variety of the
parameters (e.g., angular momentum profile) is computationally very
expensive. The purpose of the present work is to understand the nature
of the non-axisymmetric instability of the massive disks orbiting a
black hole and resulting gravitational waves emitted. For this, we
believe that the present setting is acceptable.

The initial black-hole mass is fixed to be $M_{\rm BH,0}=50M_\odot$
while a wide range of the disk mass, $M_{\rm disk}\approx
15$--$50M_\odot$, is employed.  The initial dimensionless spin of the
black hole is set to be approximately 0.8 supposing that the
very-massive stellar cores, which form a massive disk during the
collapse, should be rapidly rotating. We set the inner coordinate
radius of the disk to be close to the radius of the innermost stable
circular orbit around the rapidly spinning black hole as $r_{\rm
  in}\approx 2GM_{\rm BH,0}/c^2$, and the outer edge is varied in the
range of $r_{\rm out}/(GM_{\rm BH,0}c^{-2})=21$--53 (see
Table~\ref{table1}).  These are the plausible sizes of the dense
region of the disk formed during the collapse of rotating very-massive
stellar cores~\cite{uchida1}.  With these setups, the maximum density
of the disks is of the order of $10^{10}\,{\rm g/cm^3}$ (see
Table~\ref{table1}). Thus, the matter in the high-density region is
mildly neutron-rich. In Appendix A, we display the profiles of the
rest-mass density and angular velocity for some of the models employed
in this paper.

Numerical simulations are performed with a fixed mesh-refinement
implementation used for the neutron-star
mergers~\cite{Sekig15,Sekig16,Kyutoku18}.  We prepare 10 refinement
levels with each region composed of $241\times 241 \times 121$ uniform
grid points for $x$-$y$-$z$ (the reflection symmetry with respect to
the equatorial plane is imposed). For the finest level, the grid
spacing is $0.018 GM_{\rm BH,0}/c^2$, and the outer boundary along
each axis is located at $L\approx 1106 GM_{\rm BH,0}/c^2$ for all the
simulations. We checked that with such setting the black hole can be
accurately evolved within the fractional change of its mass and
dimensional spin of $0.1\%$ before the onset of the mass accretion
resulting from the disk instability.  


\begin{figure*}[t]
\includegraphics[width=56mm]{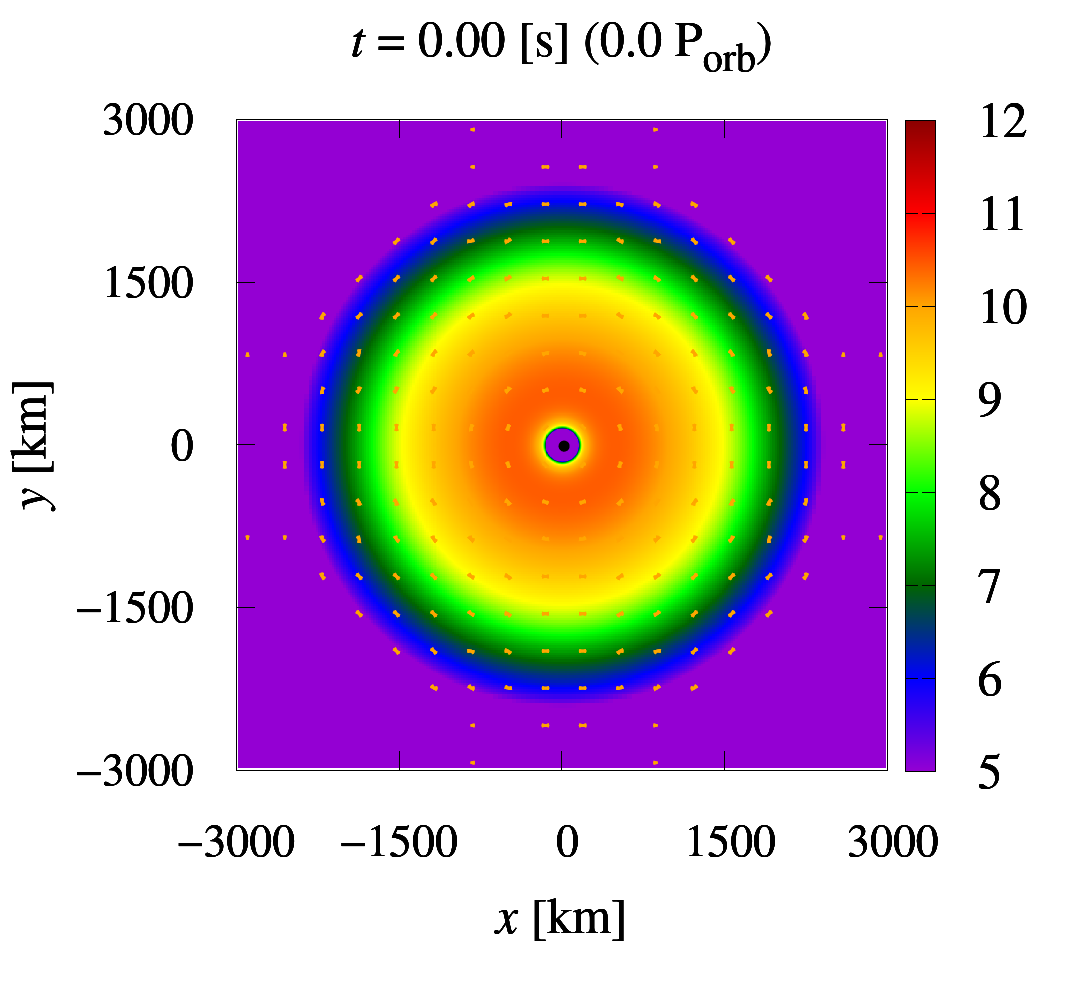}
\includegraphics[width=56mm]{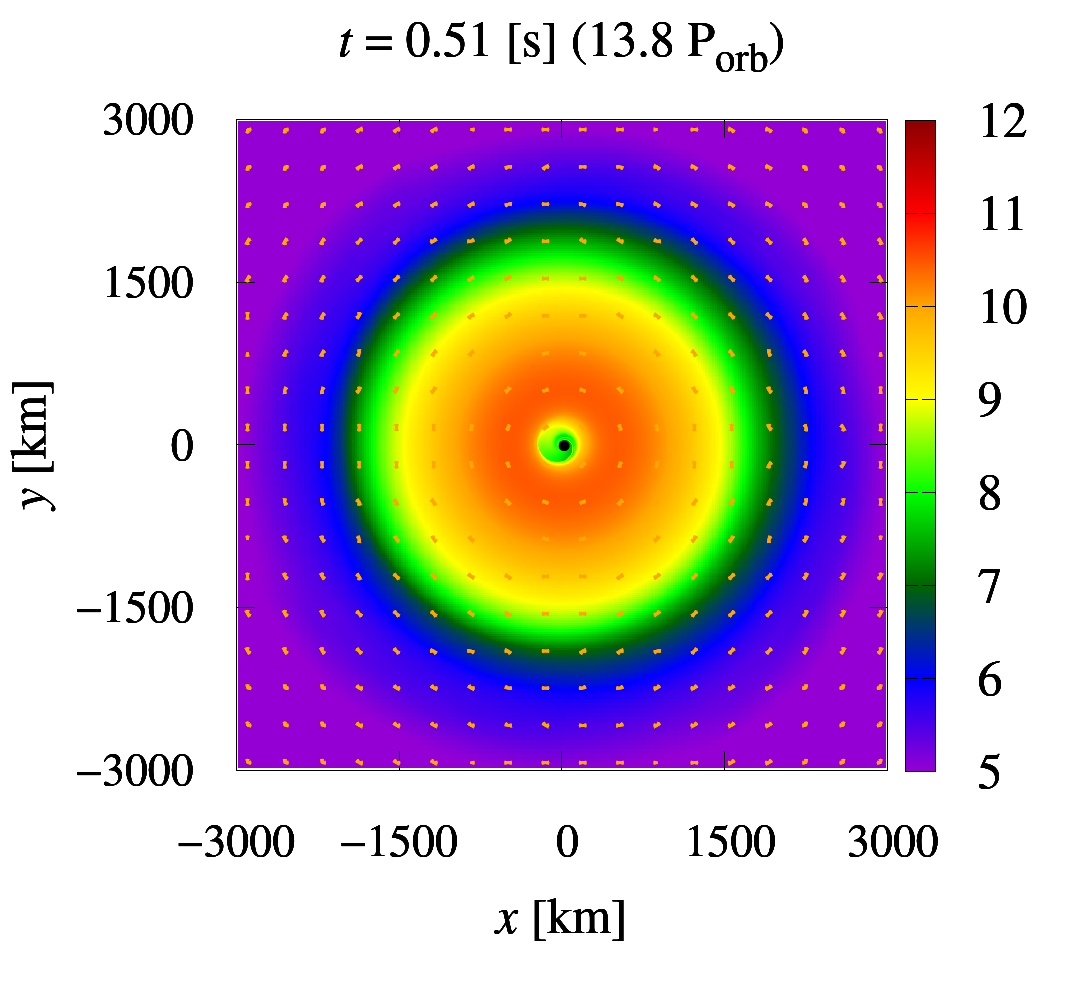} 
\includegraphics[width=56mm]{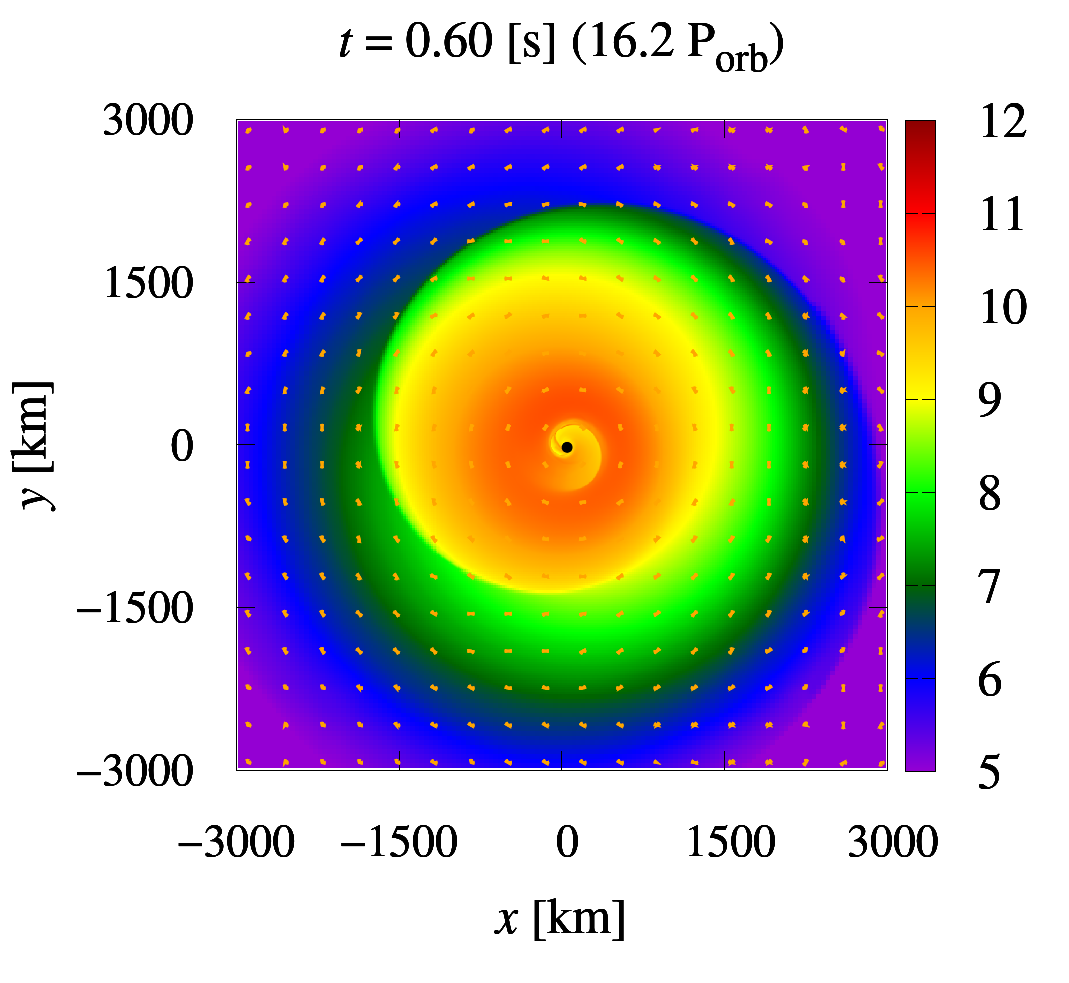}\\
\includegraphics[width=56mm]{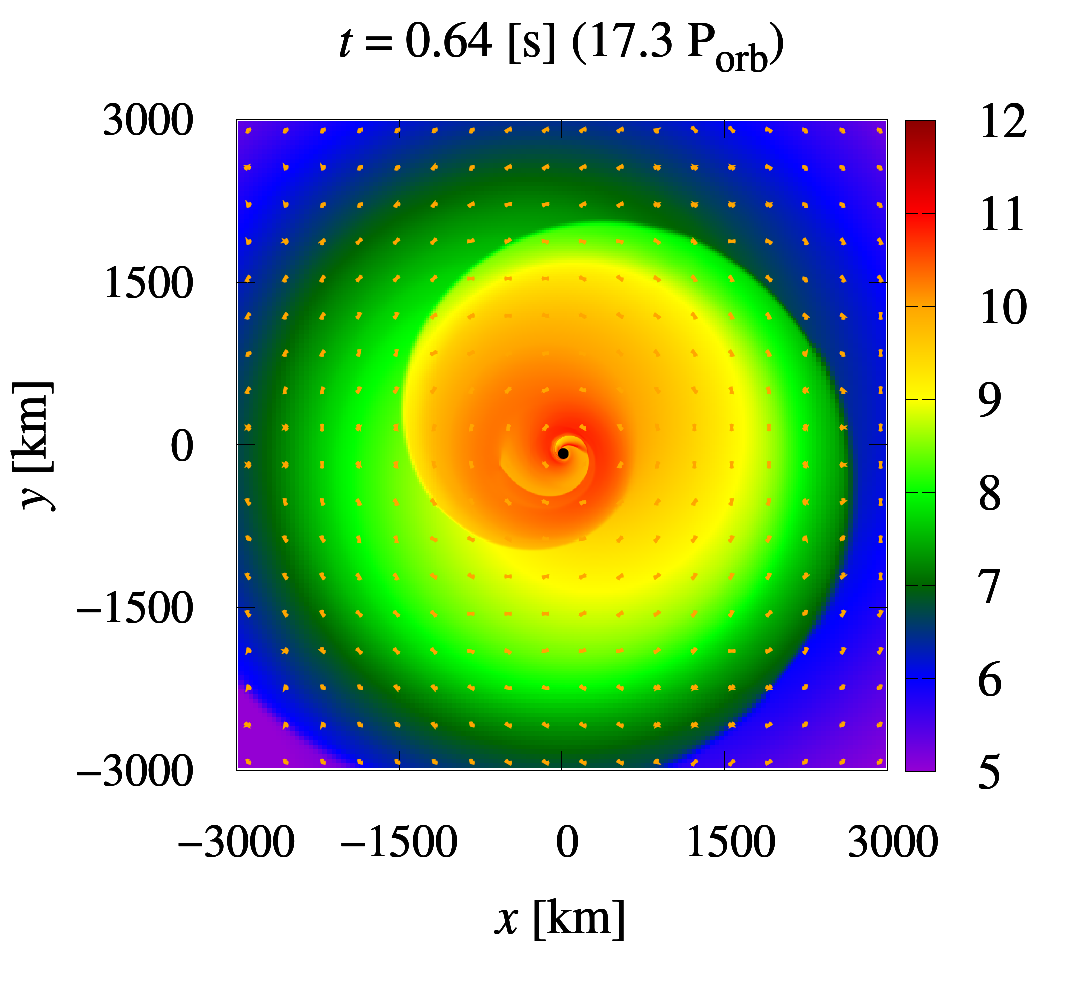} 
\includegraphics[width=56mm]{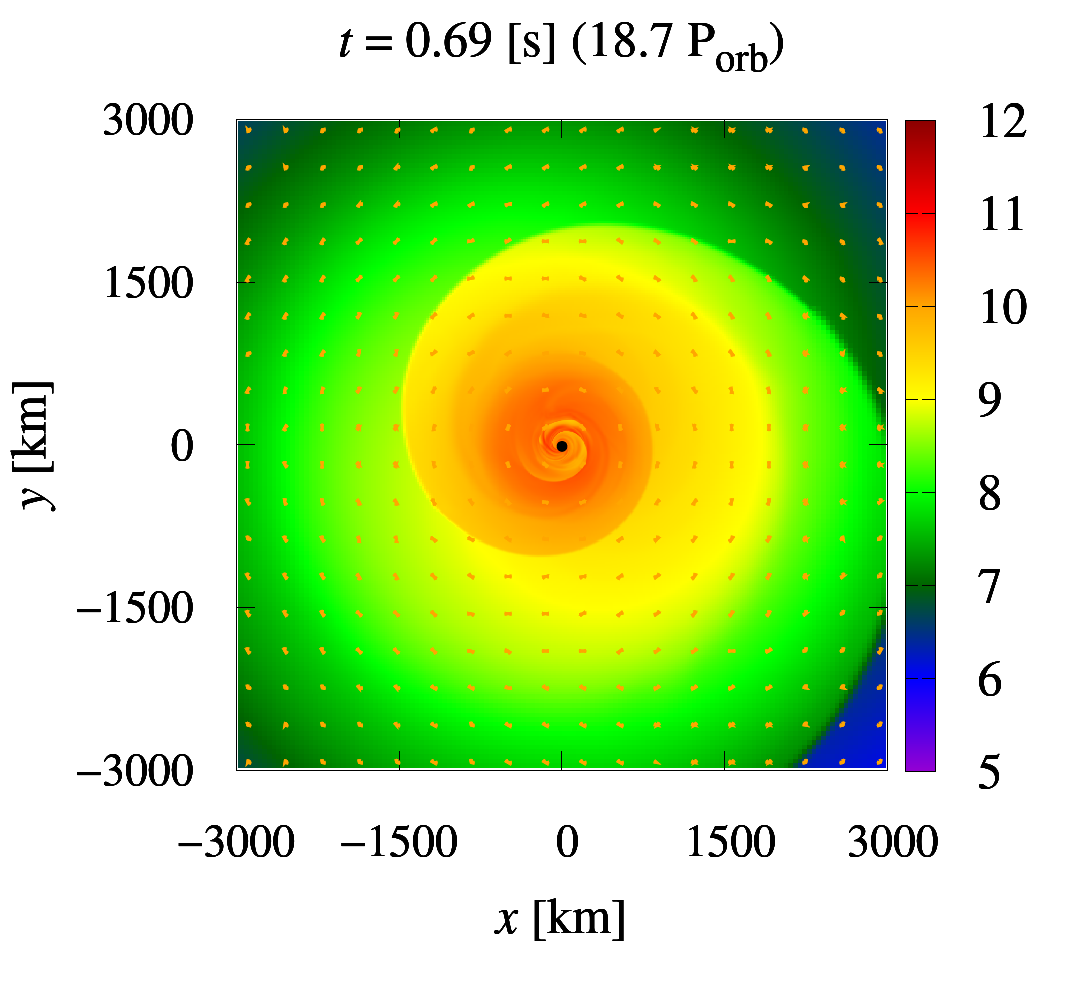} 
\includegraphics[width=56mm]{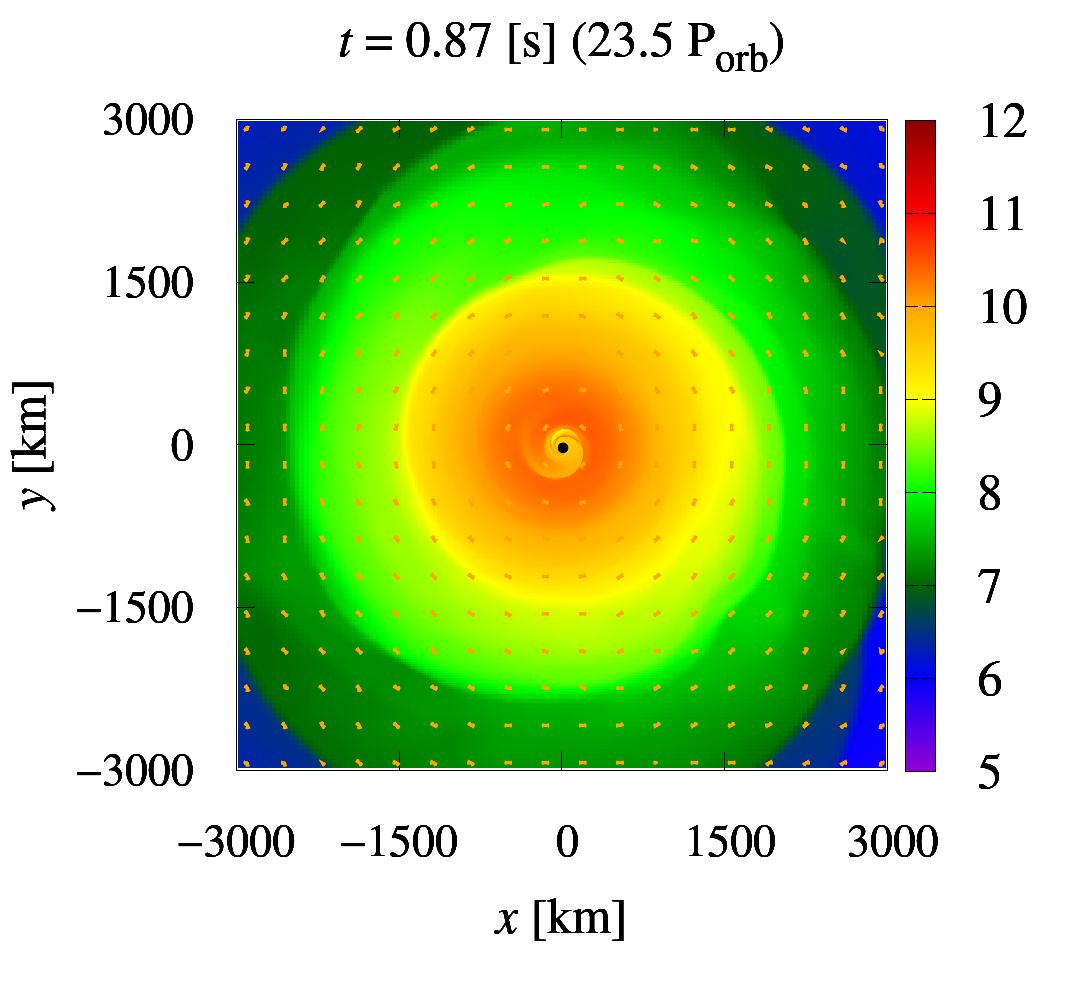} 
  \caption{Snapshots of the rest-mass density profile of the disk in
    the equatorial plane for model M14. $P_{\rm orb}$ denotes the
    initial orbital period of the disk at the density maximum (see
    Table~\ref{table1}).  The color bar shows $\log\rho$ with the unit
    of $\rho$ being ${\rm g/cm^3}$. See also
    http://www2.yukawa.kyoto-u.ac.jp/\,$\tilde{~}$masaru.shibata/BHdiskM14.mp4
    for the corresponding animation.
\label{fig1}}
\end{figure*}

\begin{figure*}[t]
(a)\includegraphics[width=84mm]{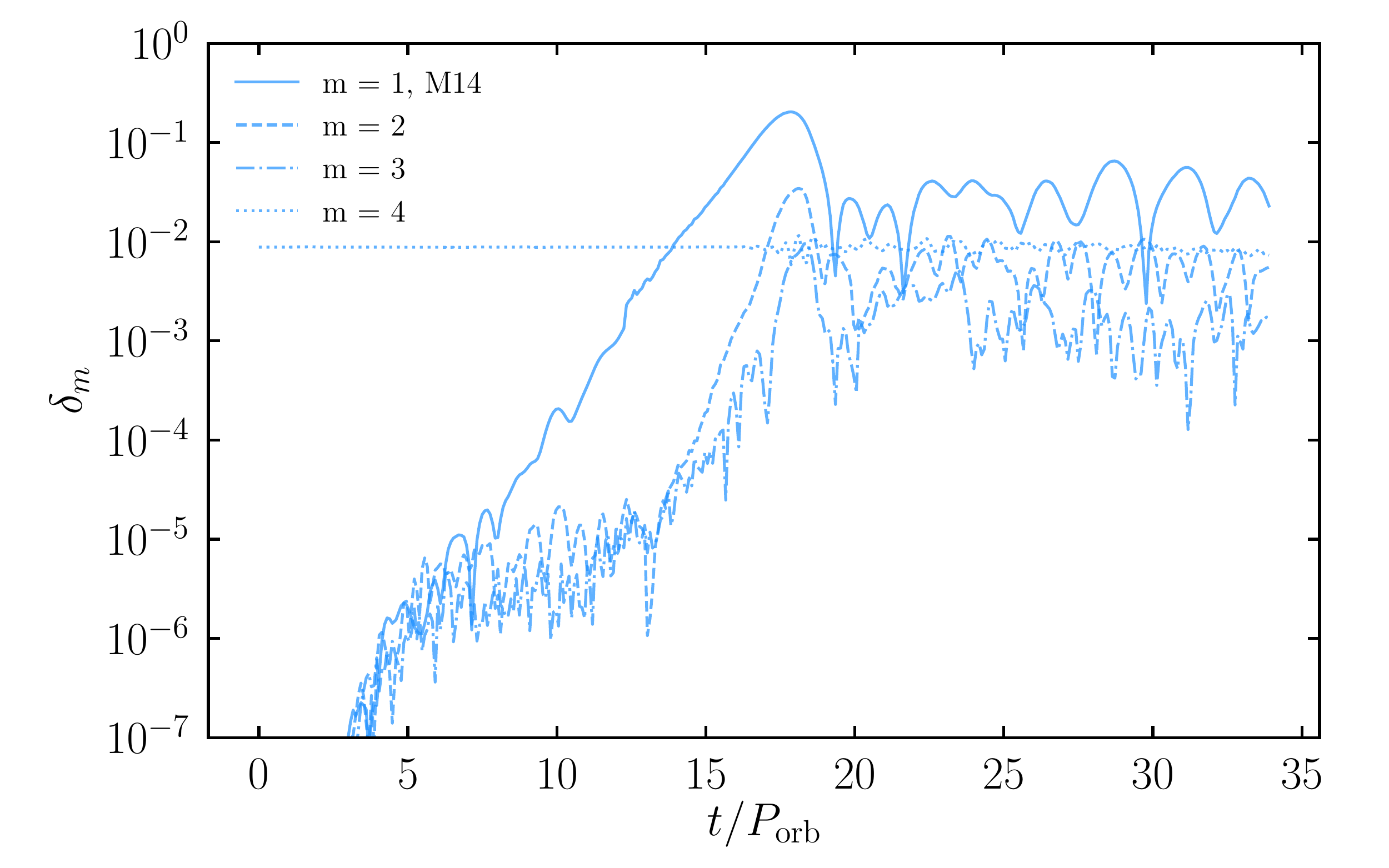}~~
(b)\includegraphics[width=84mm]{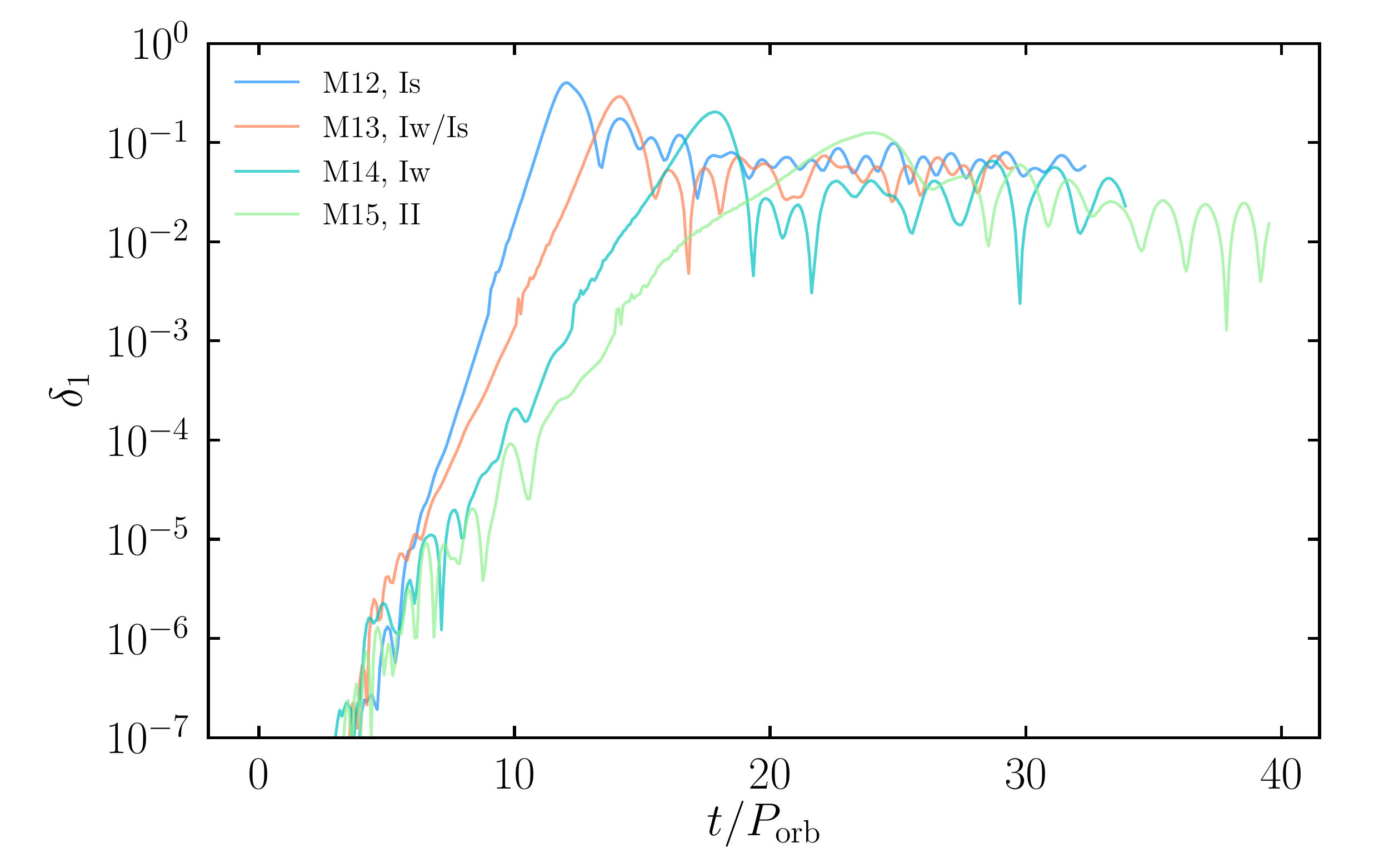}\\
(c)\includegraphics[width=84mm]{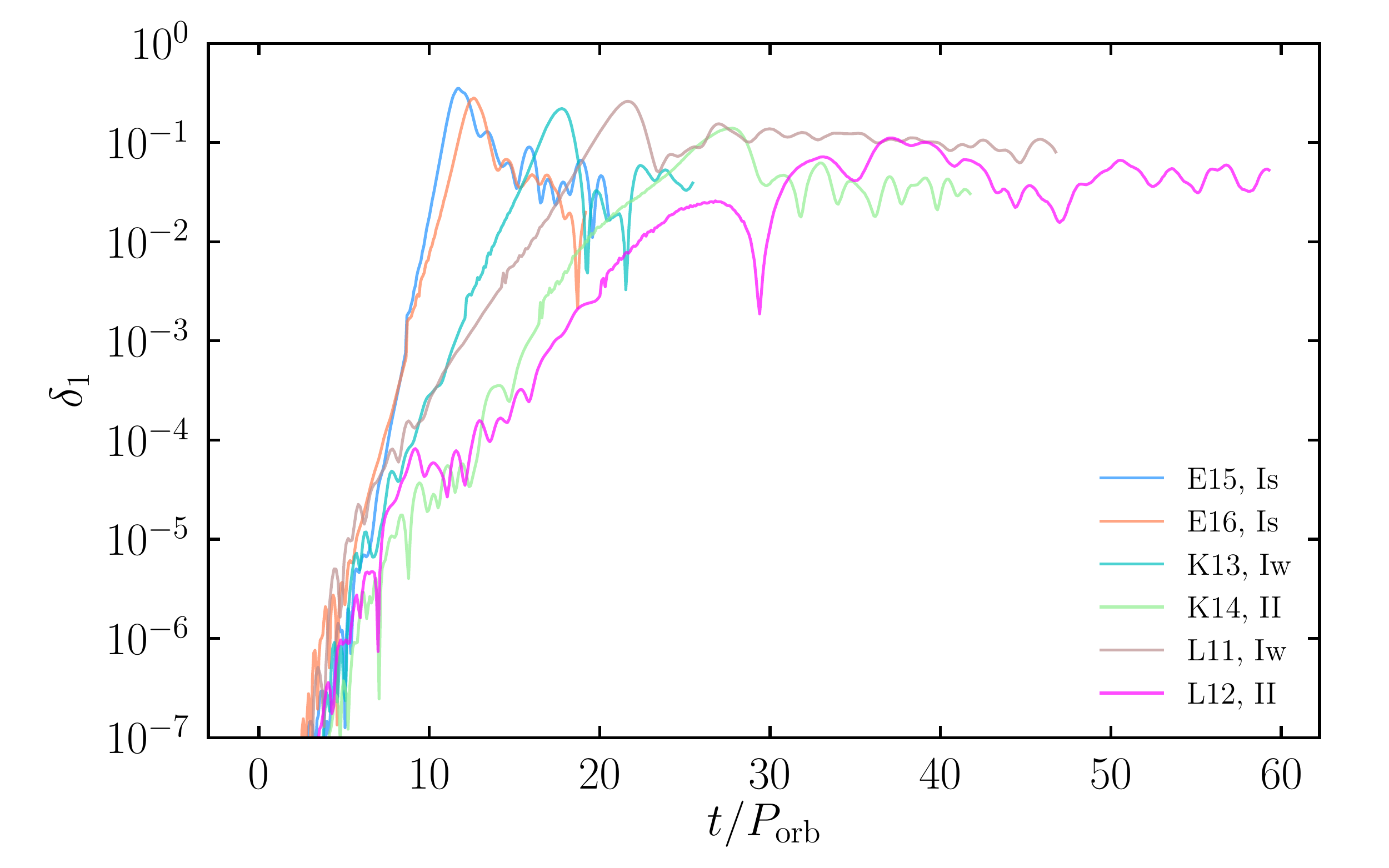}~~
(d)\includegraphics[width=84mm]{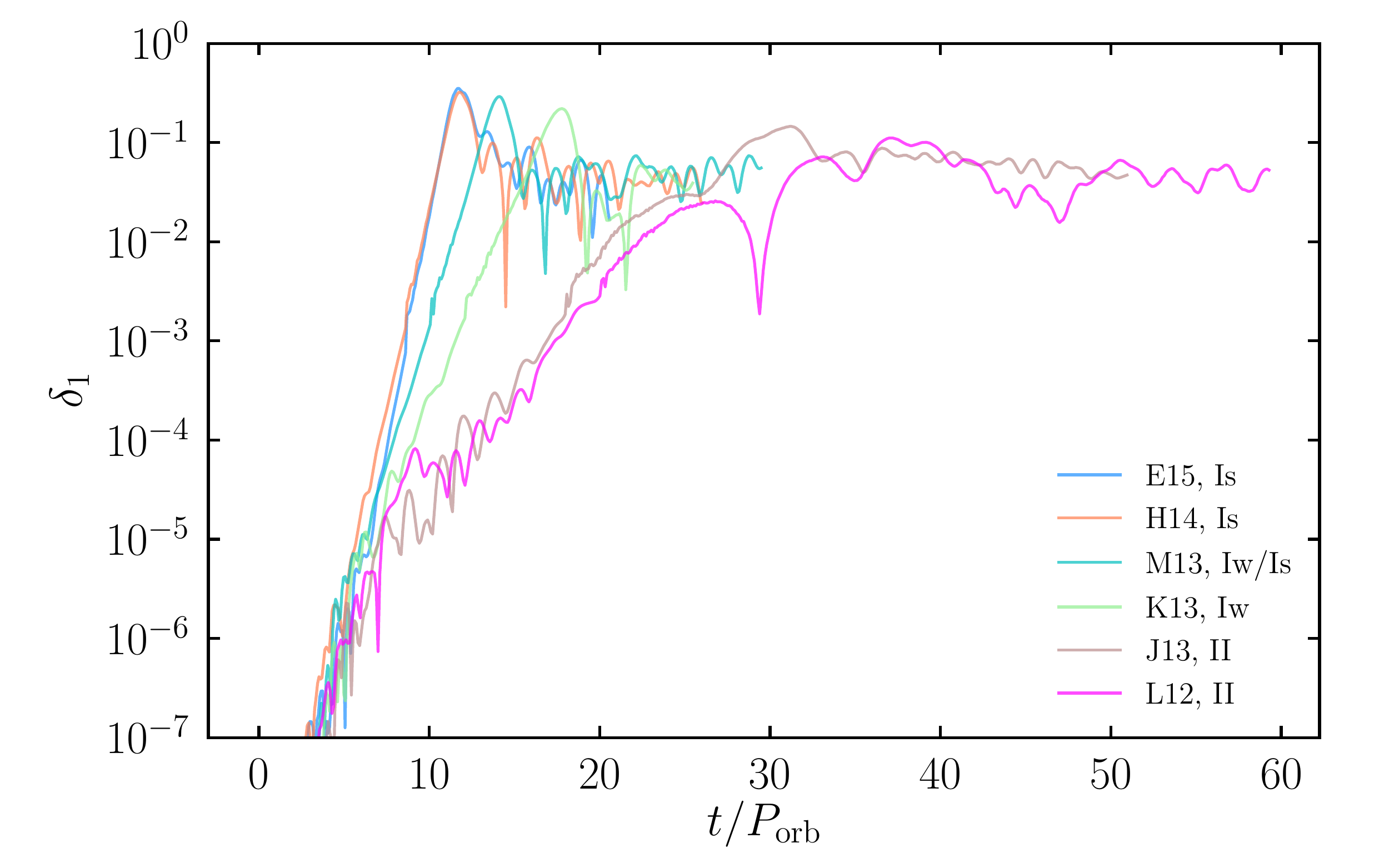}
  \caption{Mode amplitude for the density perturbation defined by
    $|C_m|/C_0$ as a function of time.  (a) Evolution of the $m=1$--4
    modes for model M14; (b) Evolution of the $m=1$ modes for models
    M12, M13, M14, and M15; (c) Evolution of the $m=1$ modes for
    models L11, L12, K13, K14, E15, and E16; (d) Evolution of the
    $m=1$ modes for models L12, J13, K13, M13, H14, and E15. The time
    is shown in units of $P_{\rm orb}$, which denotes the initial
    orbital period of the disks at the maximum rest-mass density (see
    Table~\ref{table1}). In panels (b)--(d), we describe the type of
    the waveform (see Sec.~\ref{sec3-2} for this).
\label{fig2}}
\end{figure*}

We do not add any perturbation initially, except for numerical noises.
By a small random perturbation automatically introduced in numerical
evolution, the dynamically unstable deformation of the massive disks
is enhanced for all the simulations.

\section{Evolution of disks and gravitational waves}\label{sec3}

\subsection{Evolution of dynamically unstable disks}

Irrespective of the initial conditions prepared in this paper, the
disk is dynamically unstable to the one-armed spiral-shape
deformation.  Figure~\ref{fig1} plots the snapshots of the rest-mass
density profile of the disk in the equatorial plane for model M14 (see
Table~\ref{table1} for the initial condition of this model).  It is
found that a small numerical noise introduced during the simulation
triggers the subsequent exponential growth of the one-armed
spiral-shape deformation mode. We note that in the present context
(i.e., in fully general relativistic computation), the enhanced
density perturbation in the disk enforces the black hole to slightly
shift to the higher-density side of the disk, because the center of
mass of the system should be preserved. This enhances the
gravitational attraction between the black hole and the high-density
part of the disk, and accelerates the growth of the one-armed density
perturbation. Note that this effect cannot be taken into account in
the simulations with fixed background spacetime, and thus, a fully
general relativistic simulation is needed to explore the one-armed
spiral-shape instability for high-mass disks.

After the significant growth of the spiral-shape deformation, the
angular momentum transport inside the disk, which is caused by the
gravitational torque from the non-axisymmetric density perturbation as
well as by the black-hole tidal field, is enhanced.  As a result, a
significant mass accretion onto the black hole proceeds.  During the
initial growth of the one-armed spiral-shape deformation until its
saturation, 20\%--40\% of the total rest mass of the disk falls into the
black hole. The fraction of the infall mass is larger for smaller values of
$r_{\rm out}/M_{\rm BH,0}$ irrespective of the disk mass. 
Subsequently, a weakly perturbed disk is formed, and from
such a disk, quasi-steady mass inflow to the black hole continues
(e.g., Refs.~\cite{Kiuchi11,Wessel}). For all the cases, the black
hole spins up by the mass accretion and the dimensionless spin always
becomes higher than 0.85 after a substantial fraction of the mass
falls into the black hole. In particular for the massive and compact
disk models such as M12, H14, E15, and E16, the dimensionless spin
exceeds $0.92$ due to the larger amount of the mass accretion.

Because of the presence of the non-axisymmetric structure in the disk,
the gravitational torque is exerted continuously to the matter in the
outer part of the disk, which is expanded significantly.  Even in the
present simulations of relatively short duration, we find that 1--2\%
of the total mass of the disk is ejected from the system. As a result
of the mass ejection (as well as of the mass accretion into the black
hole), the maximum density of the disk decreases (compare the first
and last panels of Fig.~\ref{fig1}).  With a longer-term simulation
taking into account the viscous/magnetohydrodynamics effects, we could
find that a more fraction of the disk matter will be ejected from the
system.  However, exploring such a long-term mass ejection process is
beyond the scope of this paper.

\begin{table*}[t]
  \caption{The growth timescale of the dynamical instability in units
    of $P_{\rm orb}$, the maximum value of $\delta_1$, and the
    signal-to-noise ratio (SNR) for gravitational waves in the
    detection at the hypothetical distance to the source of 100\,Mpc
    with the most optimistic direction assuming the detection by a
    single advanced LIGO detector with the designed sensitivity.  The
    last row describes the type of the waveform (see the text for the
    classification).}
\begin{tabular}{lcccccccccccccccc} \hline
  Model~~& ~L11~ & ~L12~ & ~J12~ & ~J13~ & ~K13~ & ~K14~ & ~M12~ & ~M13~ & ~M14~ & ~M15~ & ~H14~ & ~H15~ & ~H16~ & ~E15~ & ~E16~ & ~E17~ \\ \hline
  $\tau_{\rm dyn}/P_{\rm orb}$~~ & 1.4 & 2.3 & 1.2 & 1.7 & 1.0 & 1.5 & 0.50 & 0.74 & 1.0 & 1.4 & 0.56 & 0.81 & 1.0 & 0.52 & 0.71 & 0.82 \\ 
  $\delta_{1,{\rm max}}$ & 0.26 & 0.11 & 0.23 & 0.15 & 0.22 & 0.14 & 0.40 & 0.29 & 0.20 & 0.13& 0.32& 0.25 & 0.17 & 0.35 & 0.28 & 0.20\\
  SNR    & 16 & 6 &13 & 6 &13 & 6 & 30  & 18  & 11  & 5  & 23 & 14 & 9 & 29  & 18 &  10 \\
  Type  & Iw & II & Iw & II & Iw & II & Is & Is/Iw & Iw & II & Is & Is/Iw & Iw & Is & Is & Iw \\
  \hline
\end{tabular}
\label{table2}
\end{table*}

Figure~\ref{fig2} displays the evolution of the mode amplitude of the
density perturbation for the selected models. To define the mode
amplitude, we first extract each non-axisymmetric mode of the
density perturbation by calculating
\beq
C_m = \int \rho_* e^{i m \varphi} d^3x,
\eeq
with $m=0$--4 and $\varphi={\rm tan}^{-1}(y/x)$ and
$\rho_*$ the weighted rest-mass density~\cite{Fujiba20}. Here, the volume
integral is performed only for the outside of the apparent horizon. We
then define the normalized mode amplitude by
\beq
\delta_m={|C_m| \over C_0}. 
\eeq
In the following, we refer to $\delta_m$ as the mode amplitude. 

For all the simulations, $\delta_m$ initially increases by the
numerical noise. However, after $\delta_1$ reaches $\sim 10^{-6}$, the
one-armed spiral instability, which is developed from the initial seed
perturbation that results from the numerical noise, makes the dominant
perturbation. Thus, in the following, we focus only on the growth rate for
$\delta_1 \geq 10^{-5}$.

Figure~\ref{fig2}(a) compares the mode amplitude of $m=1$--4 for model
M14. This clearly shows that only the $m=1$ mode initially grows
exponentially with time (focusing on the plot only for the stage of
$\delta_1 \agt 10^{-5}$).  During the early exponential growth of the
$m=1$ mode, the other modes with $m \geq 2$ do not show the evidence
for the exponential growth. For the late time for which the $m=1$ mode
amplitude exceeds $10^{-2}$, we find the exponential growth for the
mode amplitudes of $m=2$ and $m=3$ but the growth of these modes
appears to result from the non-linear growth of the $m=1$ mode because
the growth rate for $m=2$ is approximately twice of that for $m=1$.
For $m=4$, the mode amplitude is approximately constant with $\delta_4
\sim 10^{-2}$. This high value of $\delta_4$ is an artifact that
results from the use of the Cartesian coordinates for the simulation.
Since this does not grow in time, no impact (and no harmful effect) is
made by it for the evolution of the system.

Figure~\ref{fig2}(b)--(d) display the $m=1$ mode amplitude,
$\delta_1$, as a function of time for several models. These show that
$\delta_1$ increases exponentially with time for the stage of
$\delta_1 \agt 10^{-5}$, for which we extract the growth timescale,
$\tau_{\rm dyn}$, by fitting $\delta_1$ in the form of $\propto
\exp(t/\tau_{\rm dyn})$.  It is found that the exponential growth
timescale for the models chosen in this paper is 0.5--2.5 times of the
initial orbital period, $P_{\rm orb}$, of the matter at the
highest-density region of the disk (cf.~Table~\ref{table2}). This
implies that the instability found here is indeed the dynamical
instability.

Figure~\ref{fig2}(b) compares the $m=1$ mode amplitude as a function
of time for models M12--M15 for which the compactness of the disk
defined by $GM_{\rm BH}/(c^2r_{\rm out})$ is different for a fixed
value of the disk mass.  In Fig.~\ref{fig2}(c) we also compare the
$m=1$ mode amplitudes between E15 and E16, between K13 and K14, and
between L11 and L12 to see the dependence of the growth timescale of
the instability on the disk compactness for a fixed value of the disk
mass.  It is always found that the growth timescale of the dynamical
instability (in units of $P_{\rm orb}$) is shorter for the more
compact disk models.  We note that if we measure the growth timescale
in units of $GM_{\rm BH,0}/c^3$, this trend is further remarkable.
Thus, we can conclude that more compact disks are more subject to the
dynamical instability. On the other hand, this result also suggests
that a sufficiently less compact disk may be dynamically stable.
Figure~\ref{fig2}(b) also shows that the maximum value of $\delta_1$
is larger for more compact disks (see also Table~\ref{table2}).

Figure~\ref{fig2}(d) compares the $m=1$ mode amplitude for models L12,
J13, K13, M13, H14, and E15 for which $c^2r_{\rm out}/(GM_{\rm BH})$
is $\sim 30$. This shows that the growth timescale of the dynamical
instability is shorter for more massive disks with a given disk
compactness. In particular for $M_{\rm disk}/M_{\rm BH,0} \agt 0.6$ the
growth timescale can be shorter than $P_{\rm orb}$.
On the other hand, this figure suggests that for a disk of mass
smaller than a critical value the disk may be dynamically
stable~\cite{Kiuchi11}. This figure also illustrates that the maximum
value of $\delta_1$ is larger for more massive disks (see also
Table~\ref{table2}).

Irrespective of the models, after $\delta_1$ reaches the maximum of
$\delta_1 \agt 0.1$, i.e., the density perturbation becomes
non-linear, the growth of the density perturbation saturates. Then,
the degree of the deformation is approximately preserved with the
density perturbation of $\delta_1 = 0.01$--0.1. This result is
qualitatively the same as, e.g., in the previous
results~\cite{Kiuchi11,Wessel}. Because the gravitational torque
resulting from the one-armed density perturbation transports the
angular momentum outward and thus the disk compactness decreases, the
value of $\delta_1$ will become smaller in the long-term evolution of
$t \gg 10P_{\rm orb}$.

In Table~\ref{table2}, we summarize the exponential growth timescale,
$\tau_{\rm dyn}$, and the maximum value of $\delta_1$, $\delta_{1,{\rm
    max}}$.  Broadly speaking, for more compact or more massive disks,
$\tau_{\rm dyn}/P_{\rm orb}$ is smaller and $\delta_{1,{\rm max}}$ is
larger.  As shown in Sec.~\ref{sec3-2}, the dependence of these
quantities on the compactness and mass of the disks is reflected in
the gravitational waveforms. We will find that three different types
of gravitational waveforms are produced by varying $r_{\rm out}$ and
the disk mass.

\begin{figure*}[t]
\includegraphics[width=80mm]{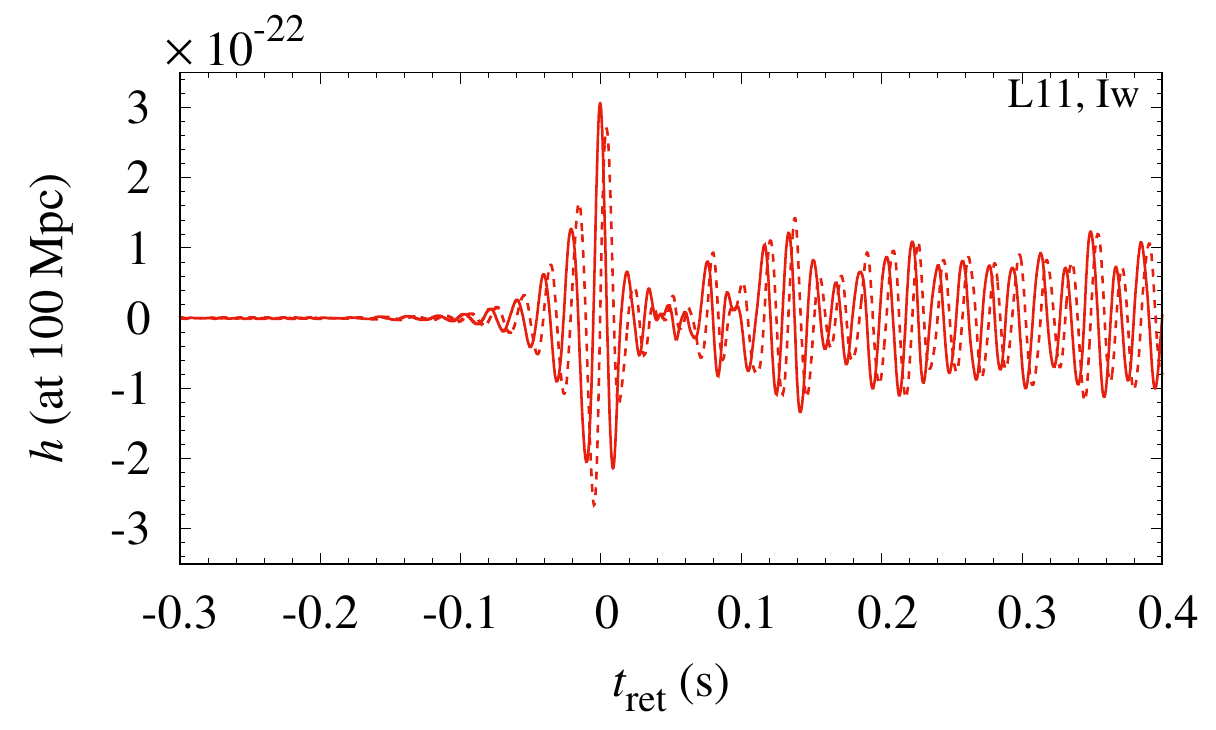}~~
\includegraphics[width=80mm]{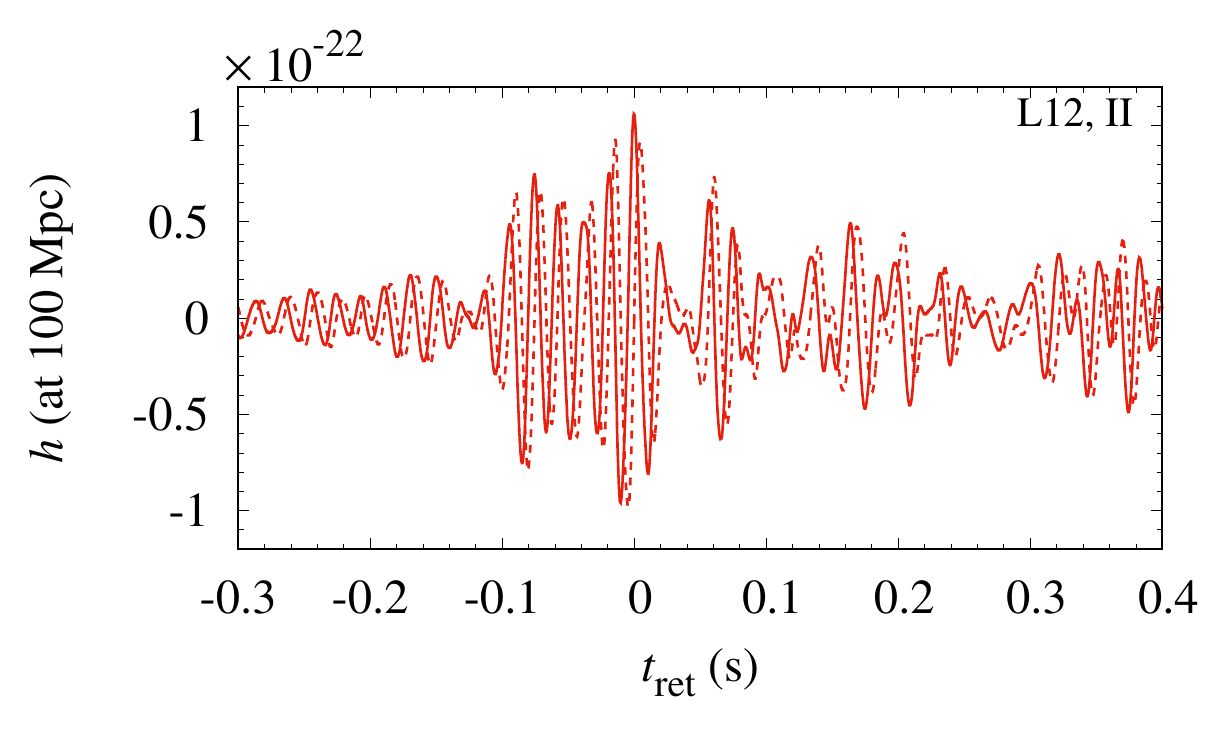}\\
\includegraphics[width=80mm]{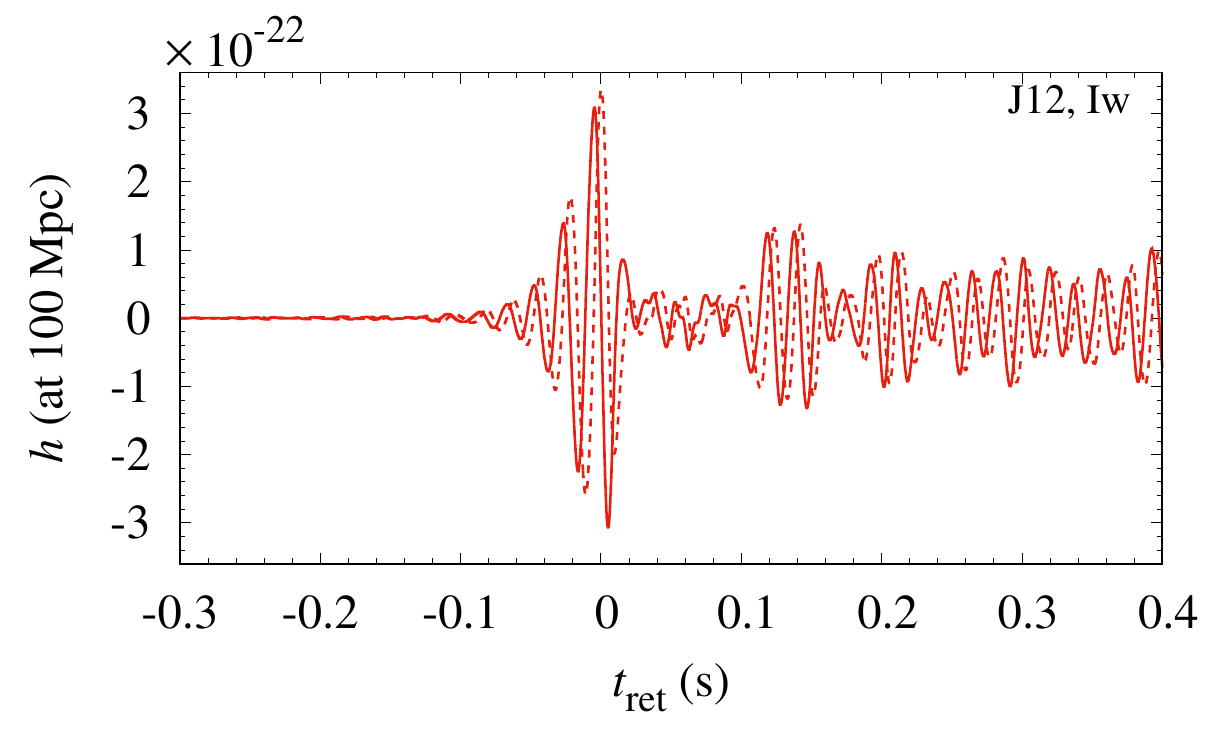}~~
\includegraphics[width=80mm]{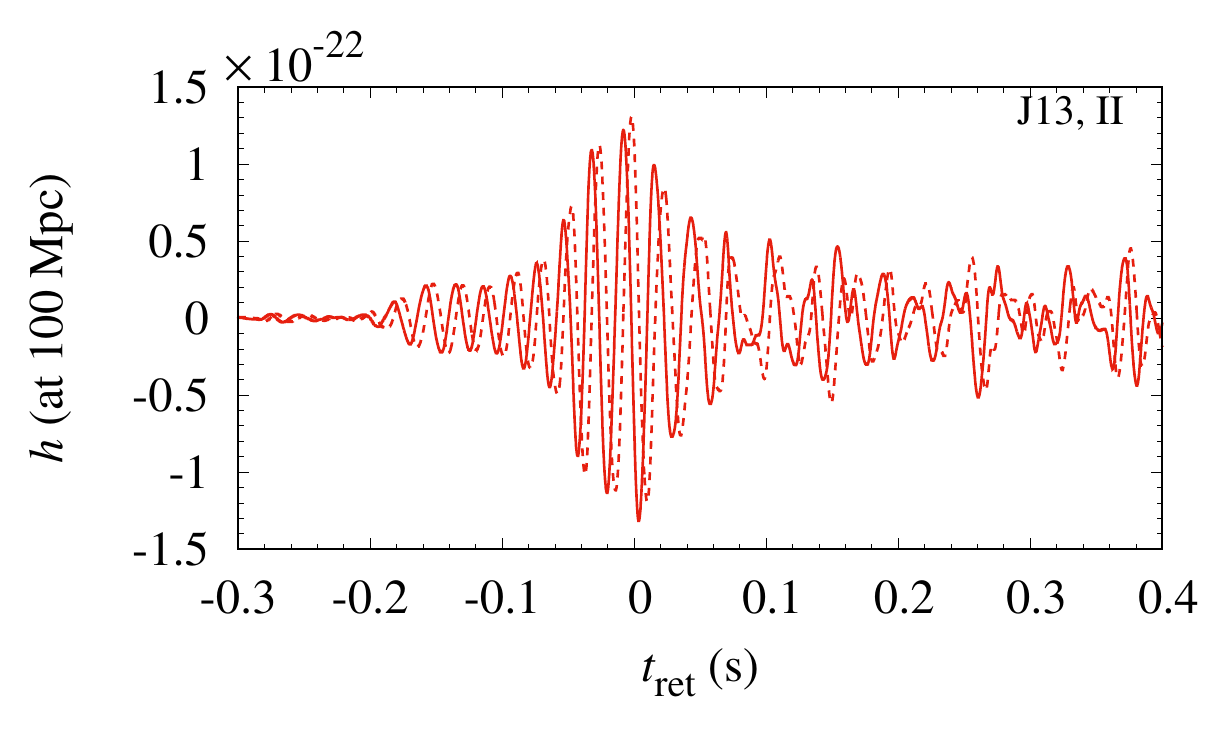}\\
\includegraphics[width=80mm]{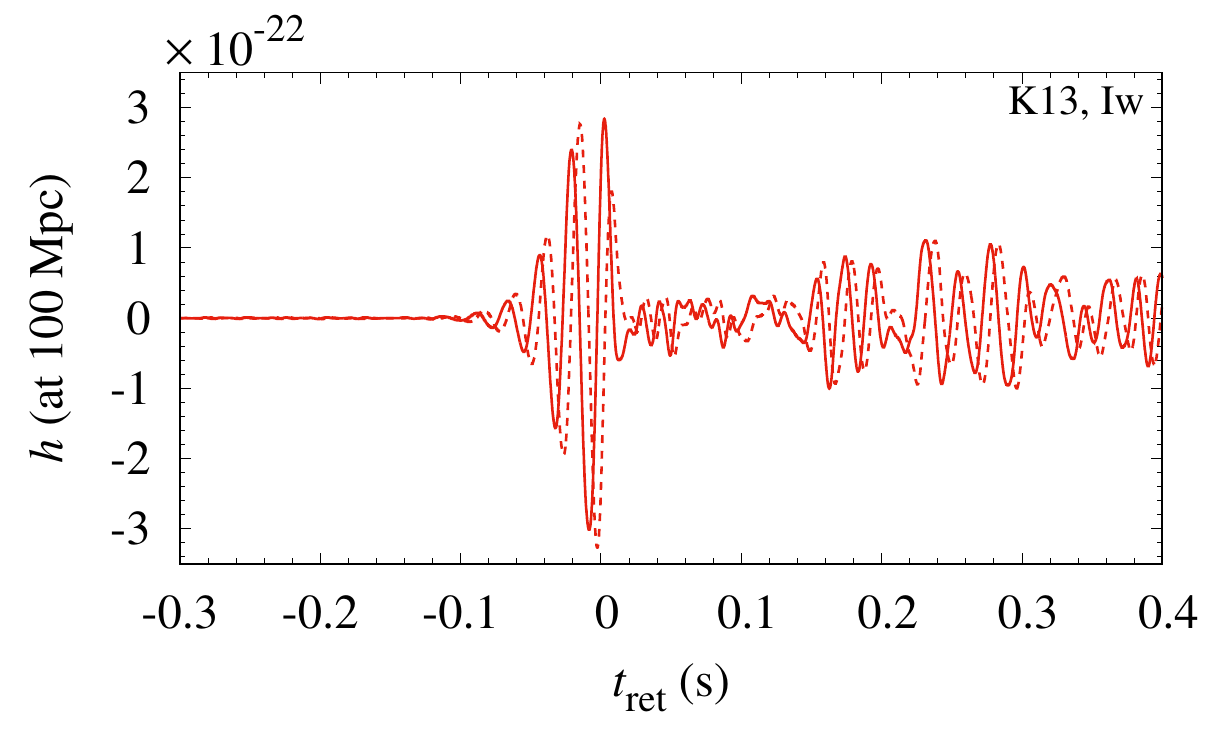}~~
\includegraphics[width=80mm]{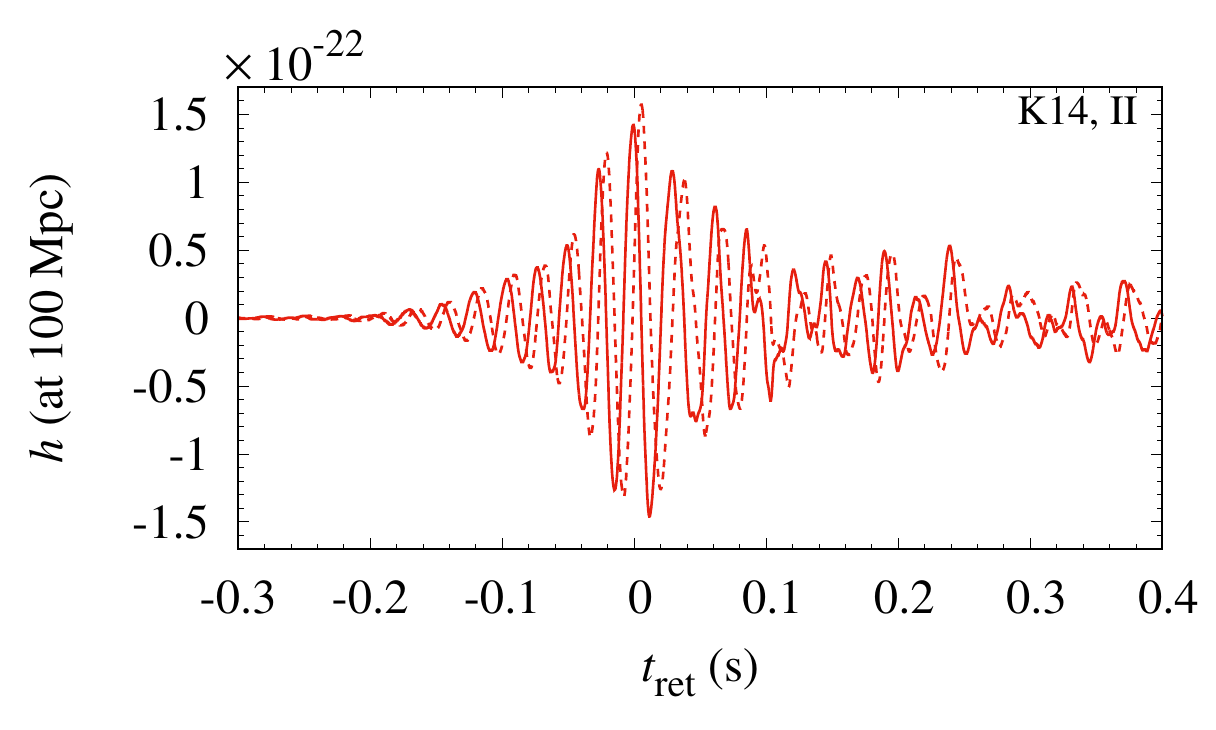}\\
\caption{Gravitational waves observed along the rotational axis at the
  hypothetical distance of 100\,Mpc for the relatively low-mass disk
  models L11, L12, J12, J13, K13, and K14.  The time axis is chosen so
  that the time at the maximum amplitude becomes approximately
  zero. The solid and dashed curves denote the plus and cross modes,
  respectively. At the upper right corner in each panel, we describe
  the type of the waveform.  We note that for each panel the scale of
  the vertical axis is different.
\label{fig3}}
\end{figure*}

\begin{figure*}[tp]
\includegraphics[width=80mm]{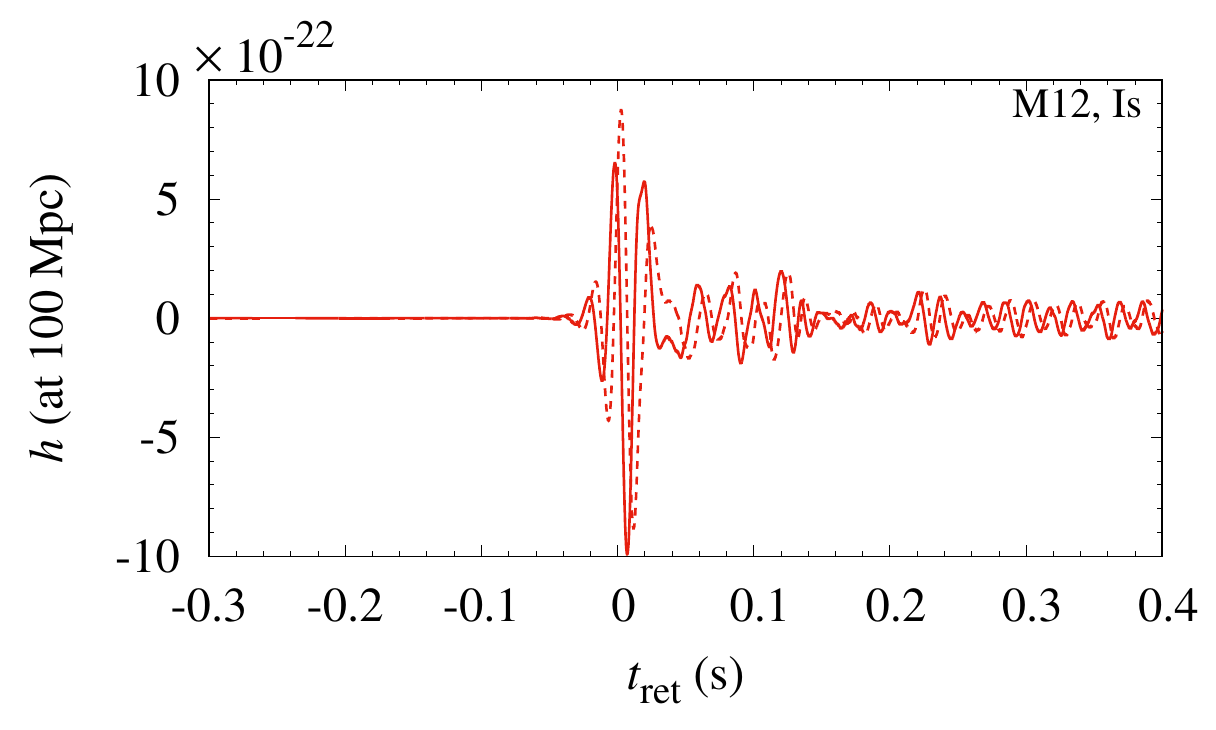}~~
\includegraphics[width=80mm]{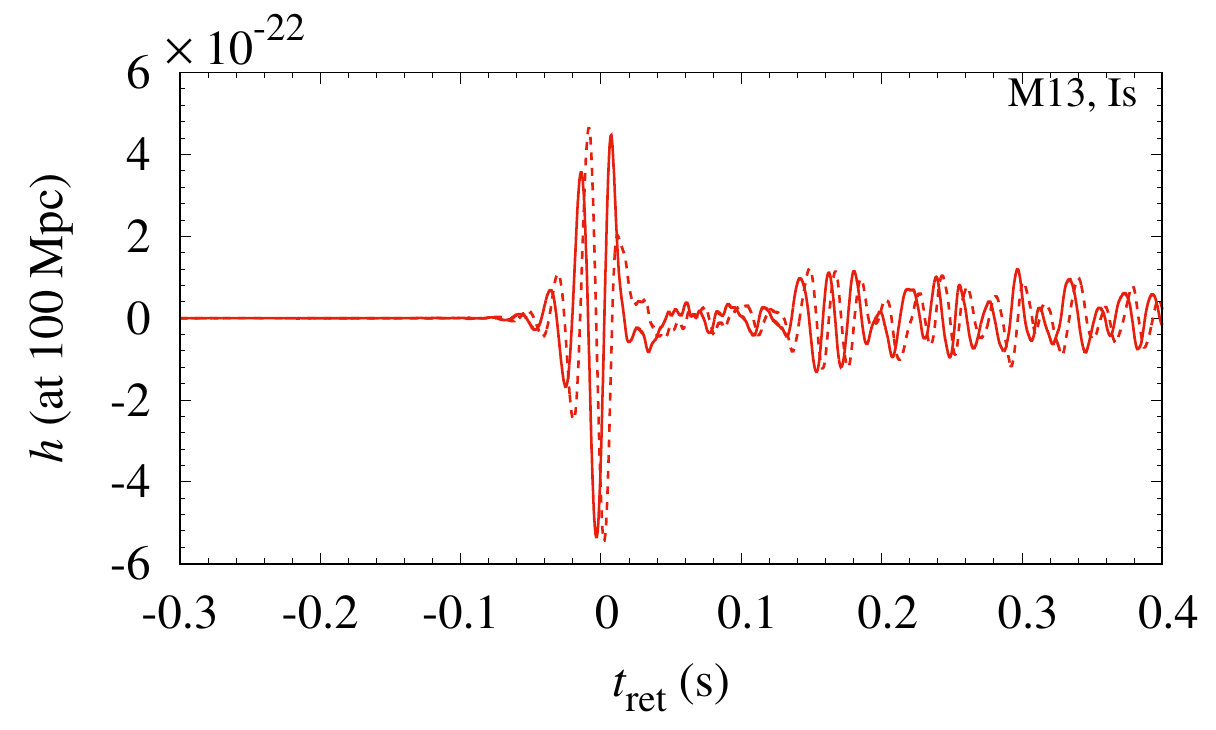}\\
\includegraphics[width=80mm]{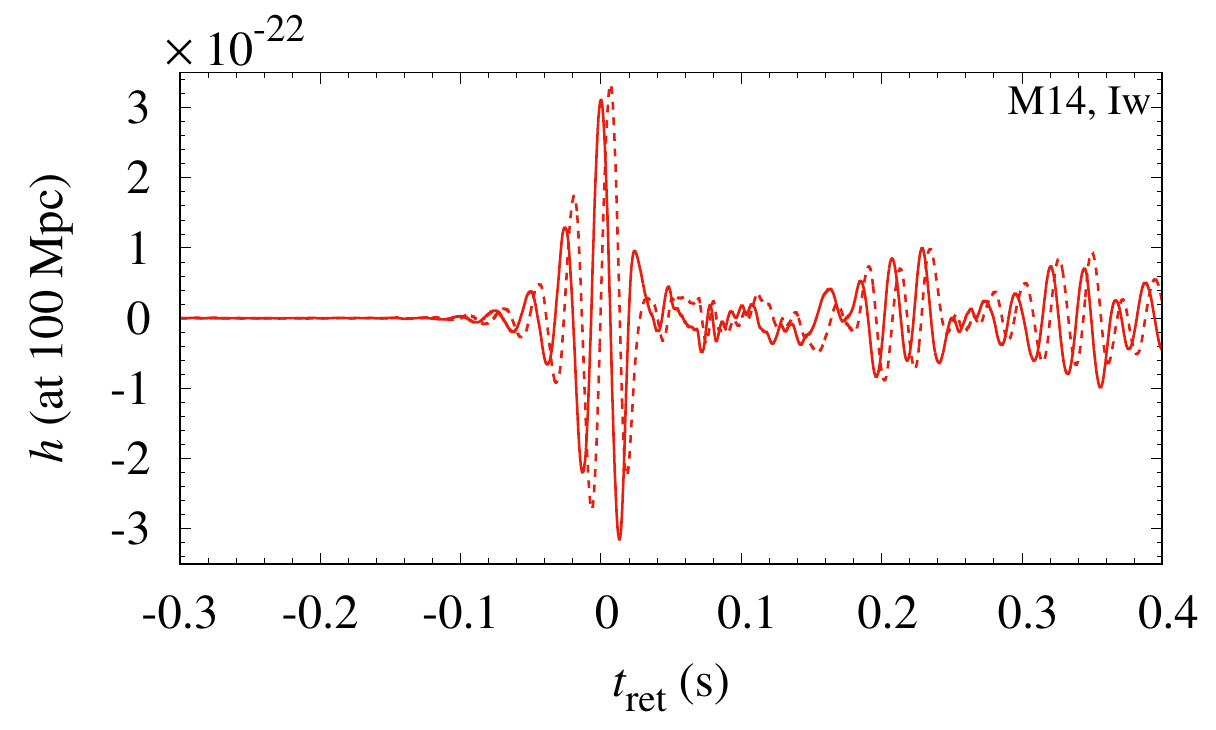}~~
\includegraphics[width=80mm]{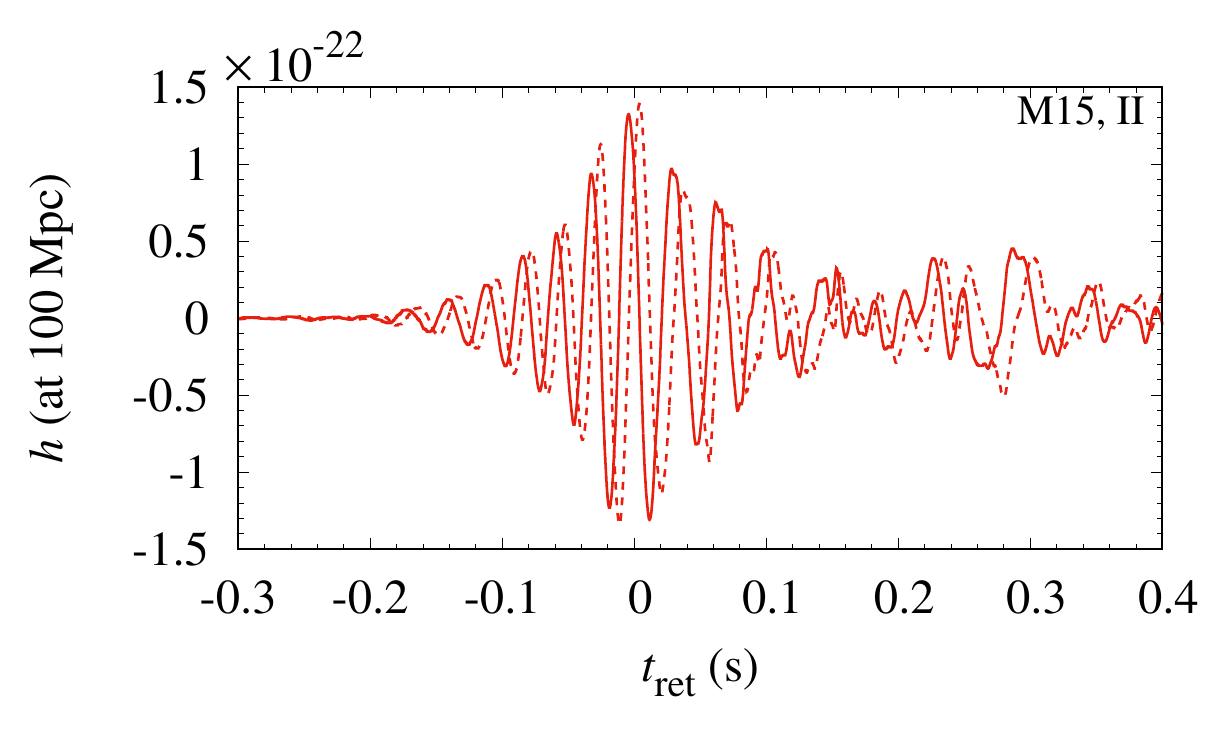}\\
\includegraphics[width=80mm]{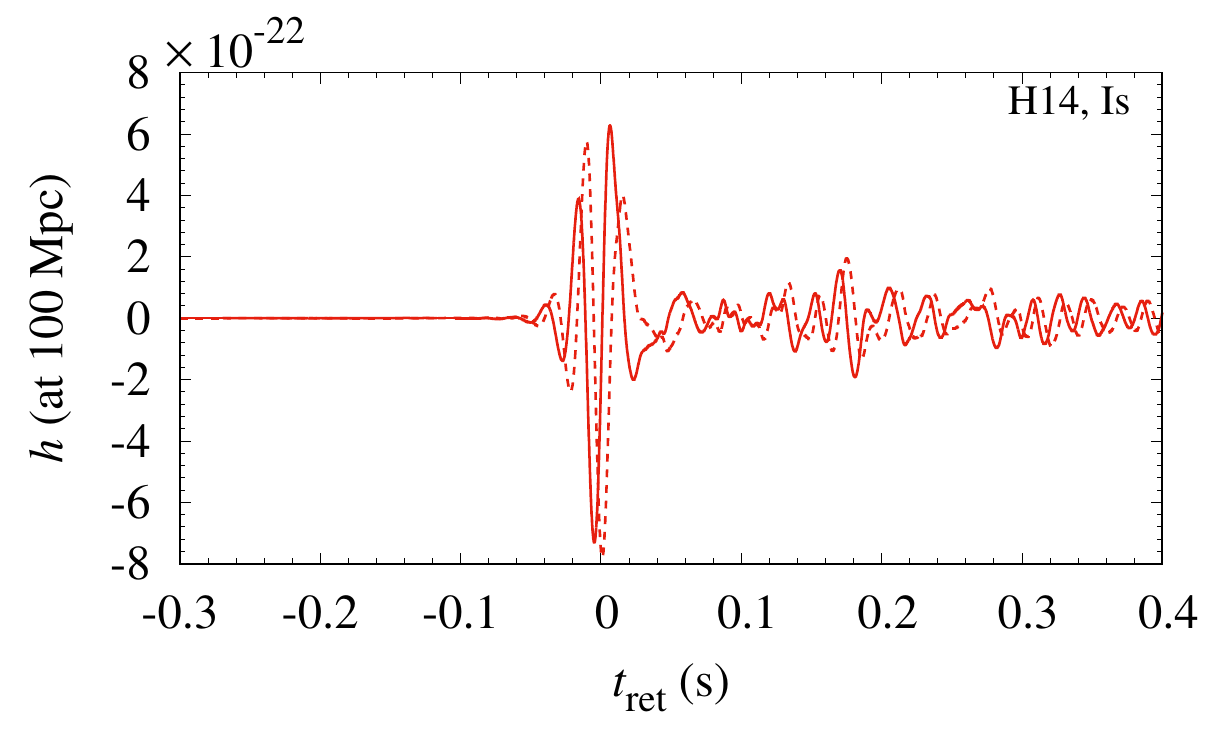}~~
\includegraphics[width=80mm]{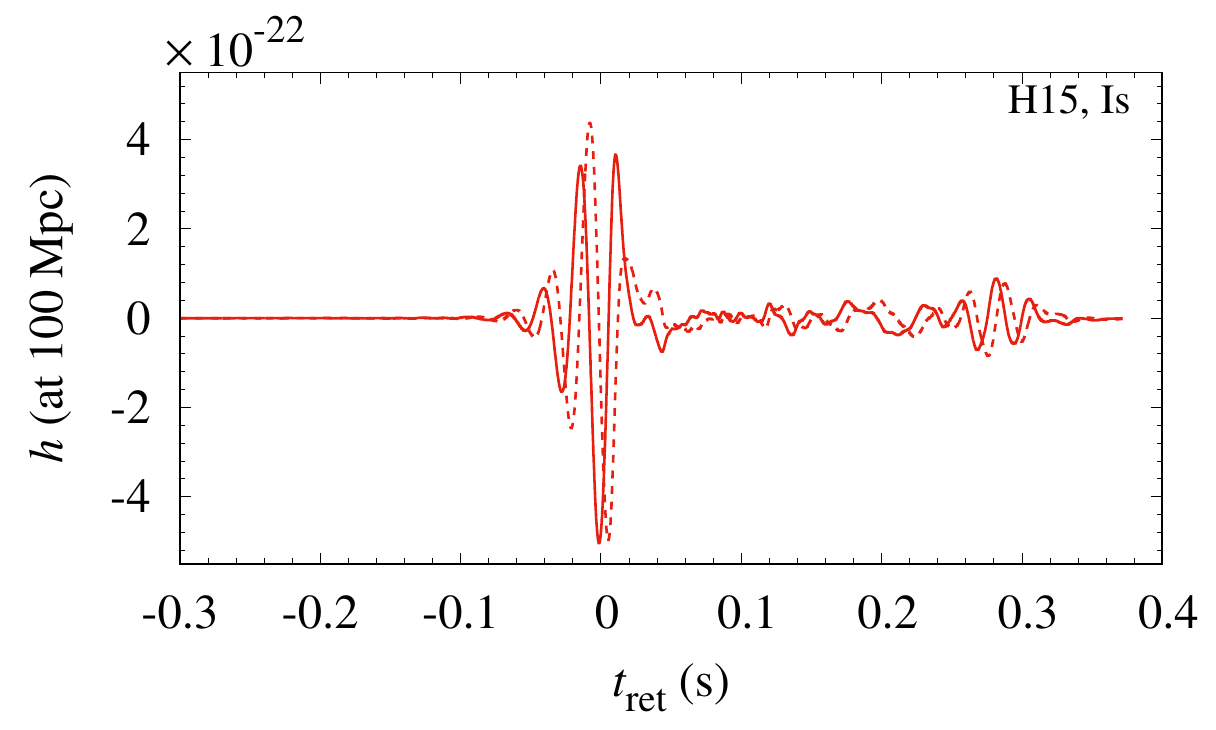}\\
\includegraphics[width=80mm]{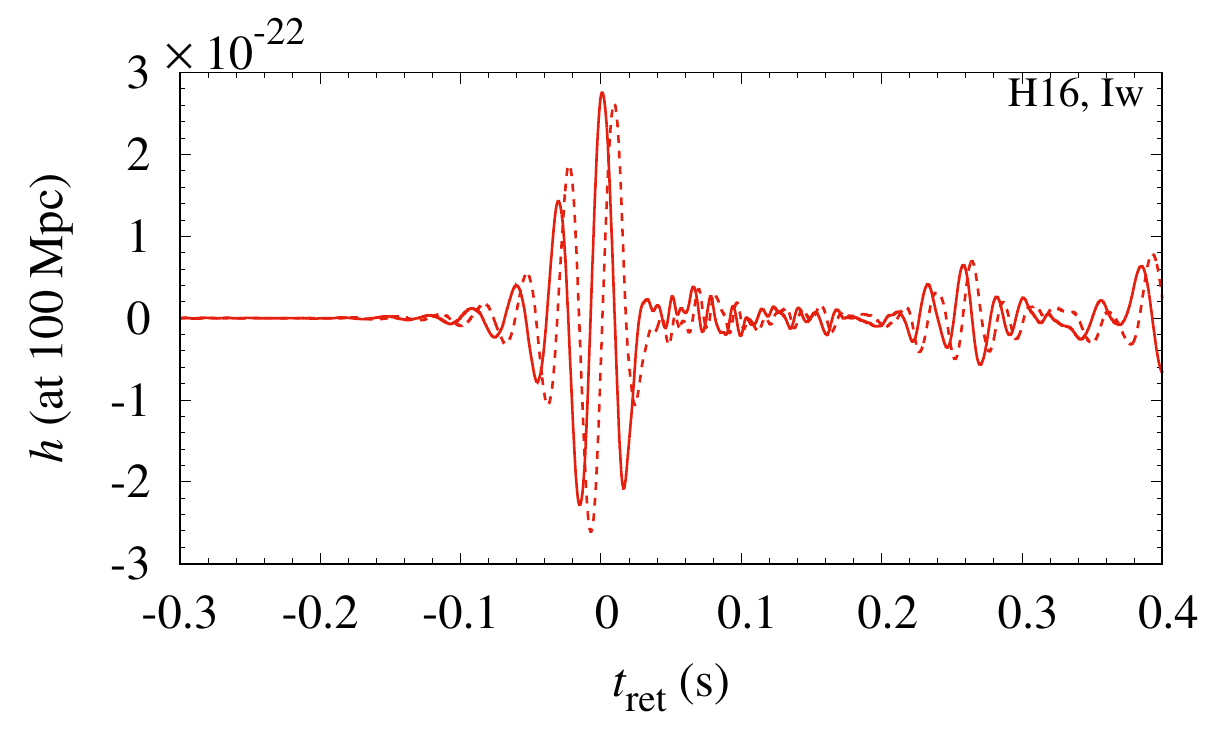}~~
\includegraphics[width=80mm]{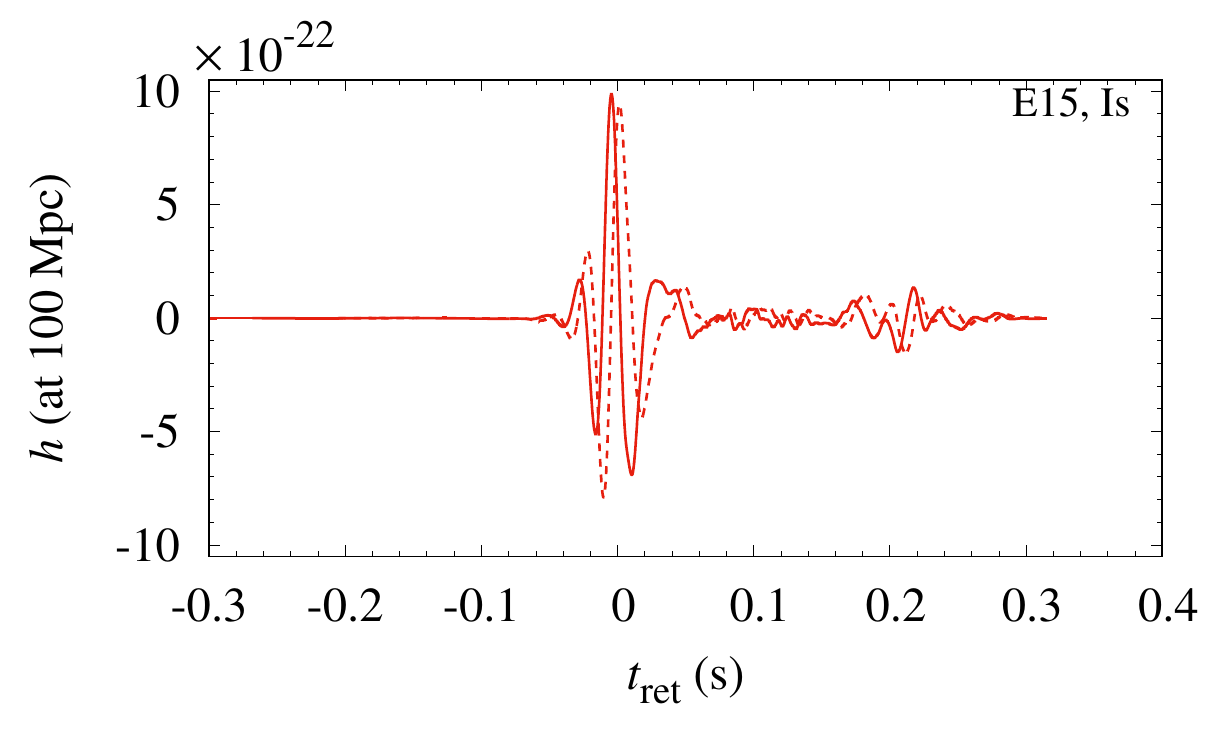}\\
\includegraphics[width=80mm]{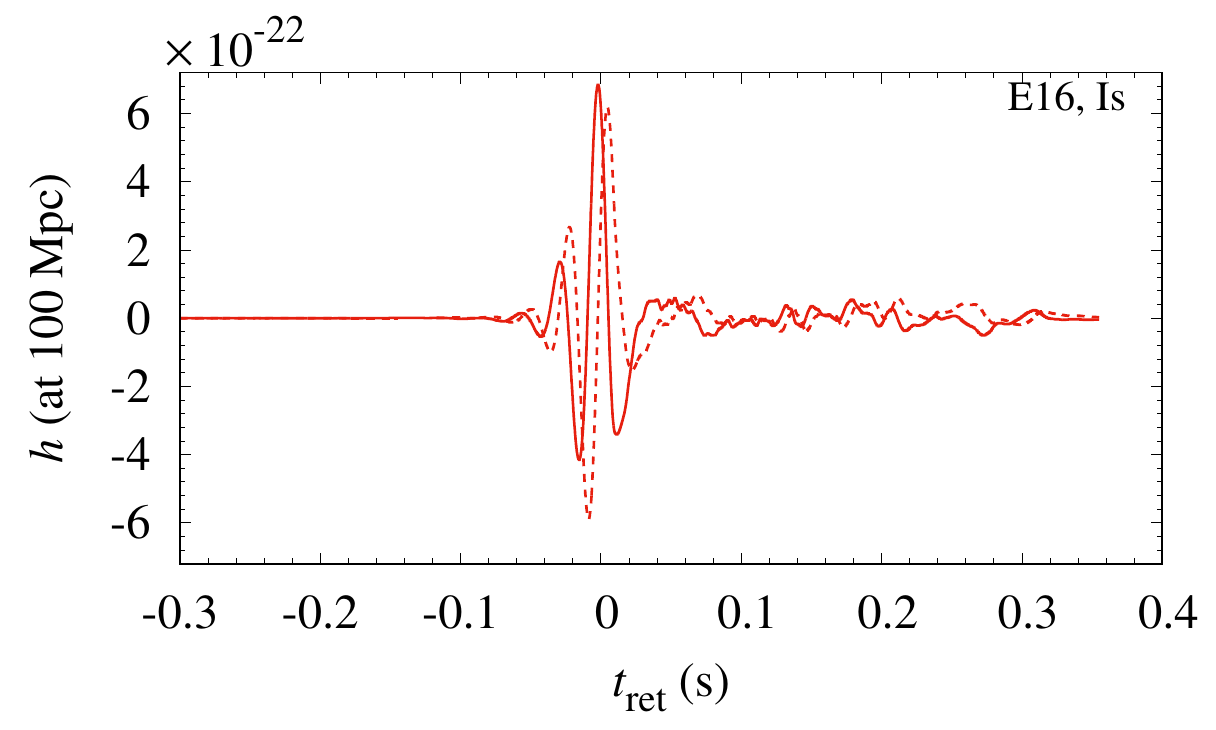}~~
\includegraphics[width=80mm]{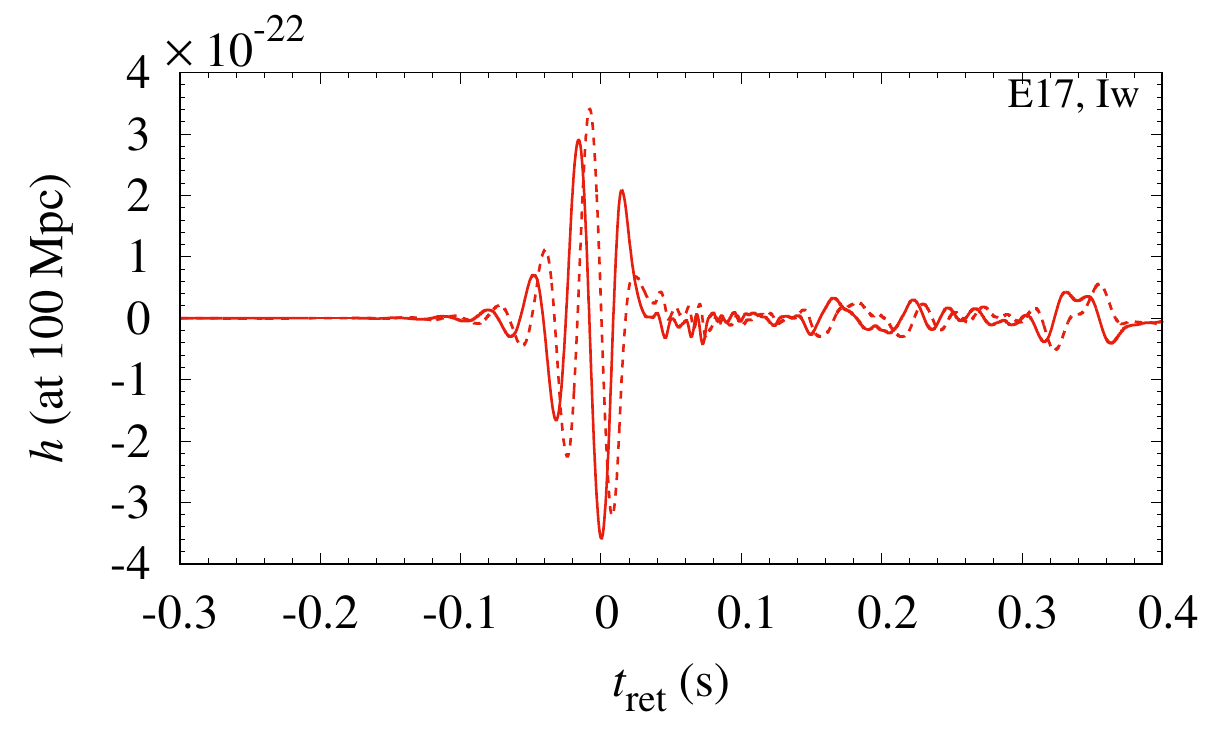}
\caption{The same as Fig.~\ref{fig3} but for the high-mass disk
  models M12, M13, M14, M15, H14, H15, H16, E15, E16, and E17. 
\label{fig4}}
\end{figure*}

Before closing this subsection, we note the following point. In the
presence of magnetic fields, the magnetorotational instability should
occur~\cite{BH98} and subsequently, a turbulence could be excited in
the disk~\cite{Hawley11,Bugli}. Then the turbulent viscosity is
enhanced, and the angular momentum in the disk would be redistributed.
If the viscous angular momentum transport efficiently works in the
disk, the growth of the one-armed spiral instability studied in this
paper should be suppressed~\cite{Bugli}.  The timescale of the viscous
angular momentum transport is
\beqn
\tau_{\rm vis}:={R^2 \over \nu},
\eeqn
where $R$ and $\nu$ denote the cylindrical radius in the disk and the
shear viscous coefficient, respectively. For the geometrically thin or
mildly thick disks, $\nu$ can be approximated as $\nu=\alpha_\nu c_s H$
where $\alpha_\nu$ is the dimensionless alpha parameter~\cite{SS73},
$c_s$ is the sound velocity, and $H$ is the scale height of the disk.
Numerical simulations for accretion disks have shown
$\alpha_\nu=O(10^{-2})$ (e.g., Ref.~\cite{Hawley11}).

Then the ratio of $\tau_{\rm vis}$ to $P_{\rm orb}$ is
\beqn
    {\tau_{\rm vis} \over P_{\rm orb}}
    \approx (2\pi \alpha_\nu)^{-1}{R^2 \Omega \over c_s H}
    \approx  (2\pi\alpha_\nu)^{-1} \left({R \over H}\right)^2,
\eeqn
where we used the approximate relation of the force balance with
respect to the vertical direction of the disk as $H \Omega \approx
c_s$. Because $R/H \agt 1.5$ for the high-density part of the disks
employed in this work (cf.~appendix A), it is found that $\tau_{\rm
  vis}/P_{\rm orb} \agt 7$ as long as $\alpha_\nu$ is not unusually
large, i.e., $\alpha_\nu \alt 0.05$. Thus, for the massive and compact
disks of $\tau_{\rm dyn}\alt 2.5P_{\rm orb}$ which we study in this
paper, the timescale of the viscous angular momentum transport is much
longer than the growth timescale of the one-armed spiral-shape
instability. This is in particular the case for the models that result
in the type I waveforms which we are most interested in (see
Sec~\ref{sec3-2}). Therefore, the magnetic/viscous effects could be
safely neglected for the models employed in this paper.  However, as
shown in Ref.~\cite{Bugli}, the magnetic/viscous effects are likely to
play an important role for low-mass or or less compact disks for which
$\tau_{\rm dyn} \agt 10P_{\rm orb}$.

\subsection{Gravitational waves}\label{sec3-2}

\begin{figure}[t]
  \includegraphics[width=86mm]{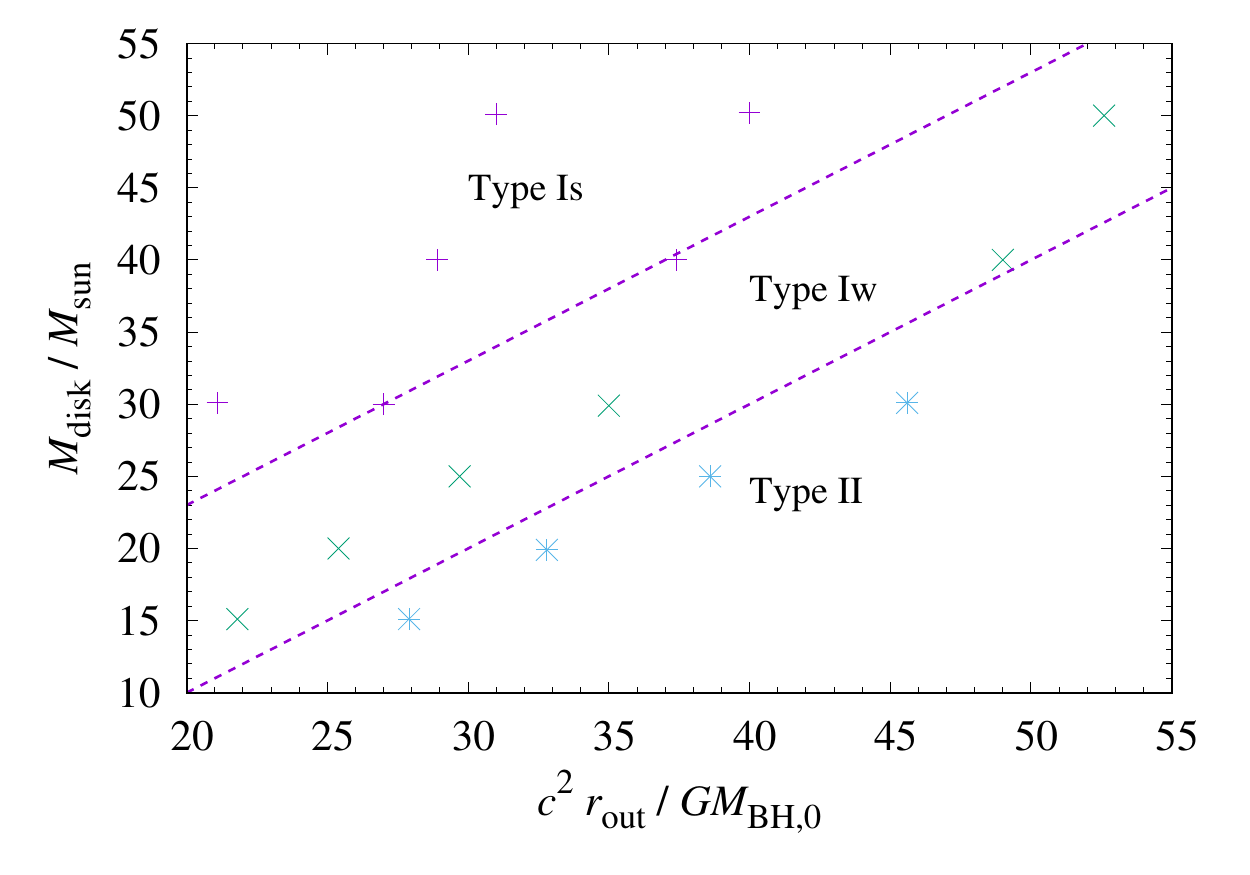}
\vspace{-5mm}
\caption{The types of the gravitational waveforms are summarized in
  the plane of $c^2r_{\rm out}/(GM_{\rm BH,0})$ and $M_{\rm
    disk}/M_\odot$. Note that the vertical axis may be interpreted as
  the mass ratio of $M_{\rm disk}/M_{\rm BH,0} (\times 50)$.  The
  upper and lower dashed lines are $M_{\rm disk}/M_\odot =3 +
  c^2r_{\rm out}/(GM_{\rm BH,0})$ and $M_{\rm disk}/M_\odot =-10 +
  c^2r_{\rm out}/(GM_{\rm BH,0})$, respectively.
\label{fig5}}
\end{figure}

\begin{figure*}[t]
(a)\includegraphics[width=84mm]{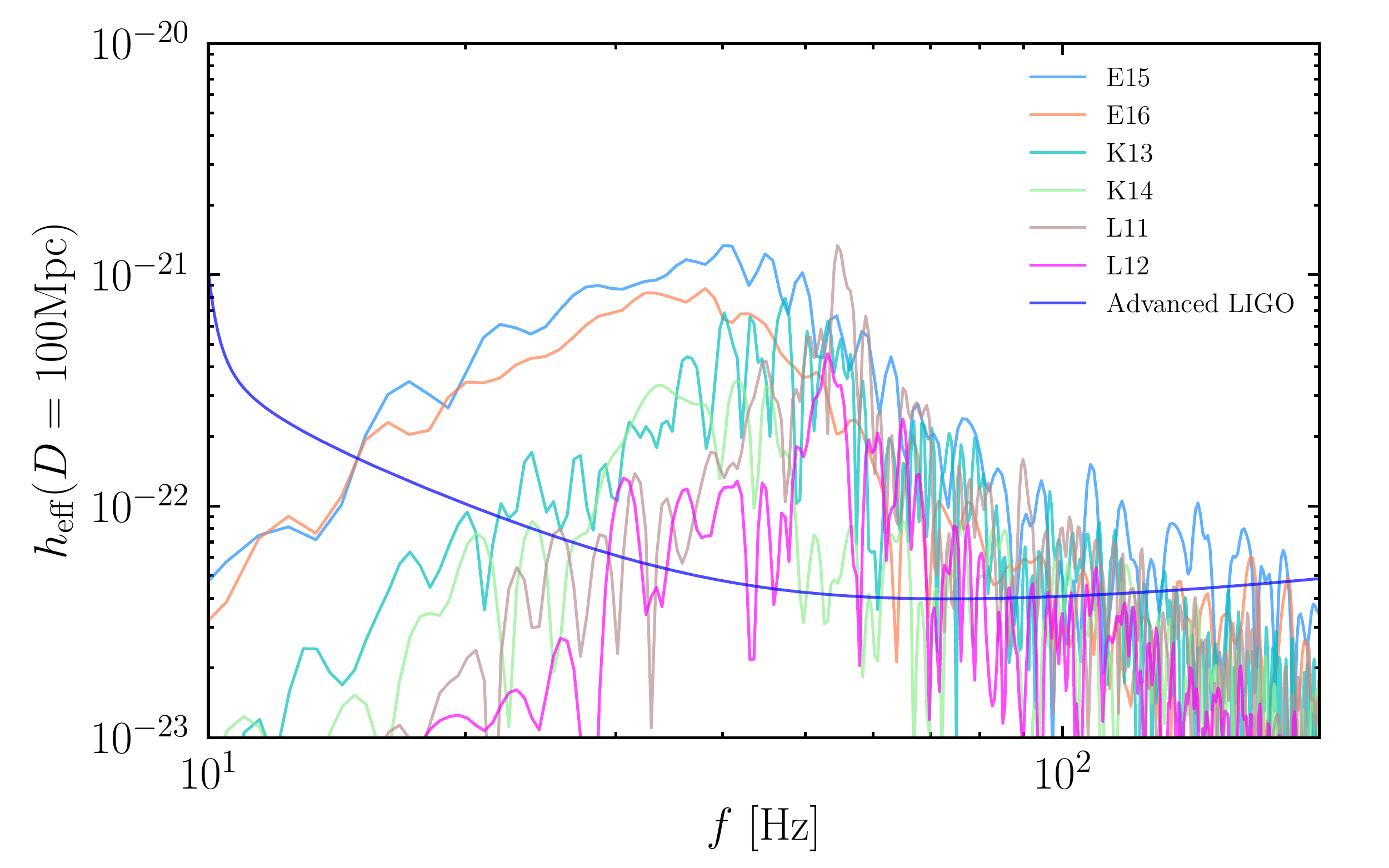} 
(b)\includegraphics[width=84mm]{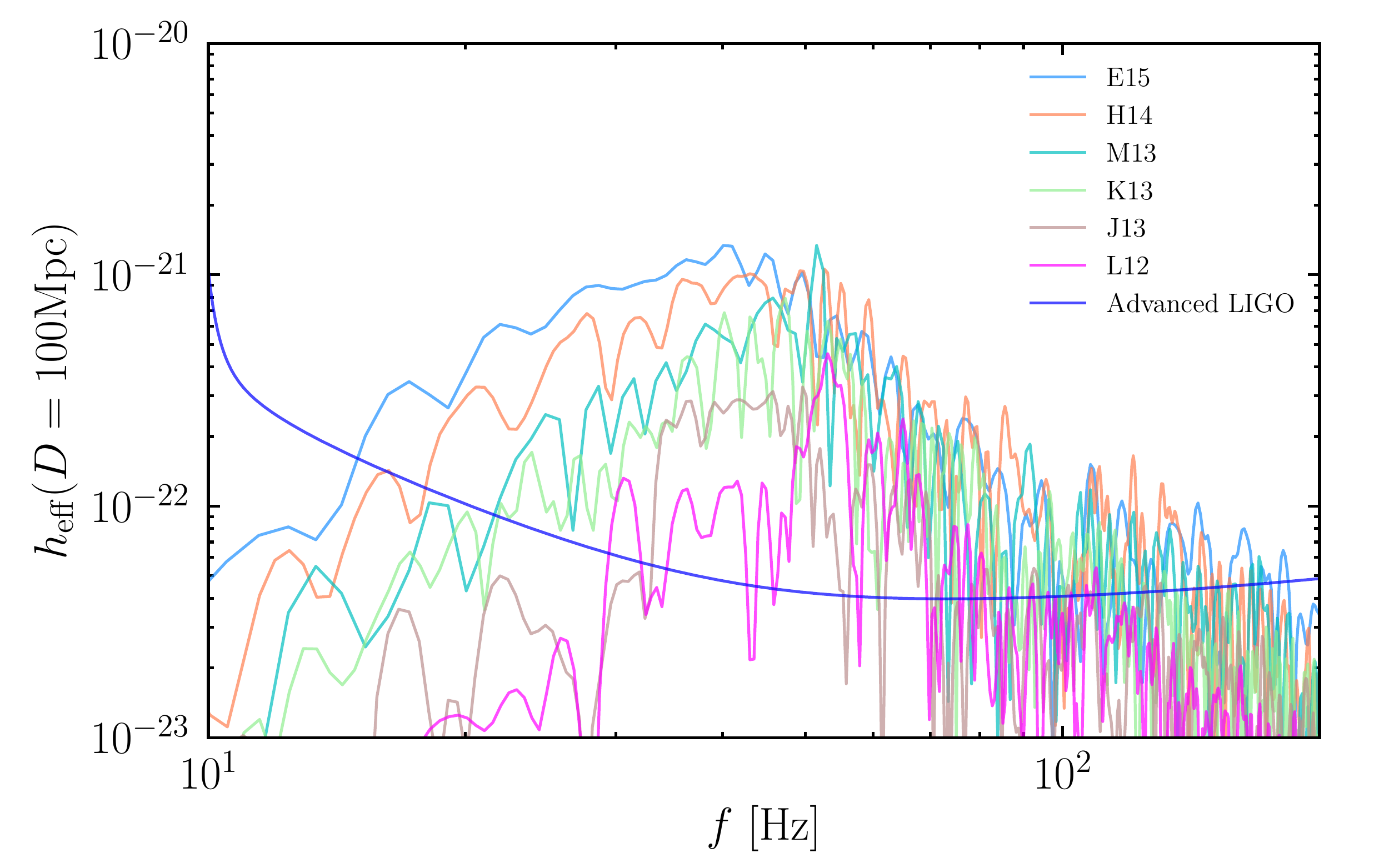} 
  \caption{Effective amplitude of gravitational waves as a
    function of the frequency (a) for models
    L11, L12, K13, K14, E15, and E16; (b) for models L12, J13, K13,
    M13, H14, and E15 for which $c^2r_{\rm out}/(GM_{\rm BH,0}) \sim
    30$.  The designed sensitivity of advanced LIGO is taken from
    https:\slash\slash{}dcc.ligo.org\slash{}cgi-bin\slash{}DocDB\slash{}ShowDocument?docid=2974.
\label{fig6}}
\end{figure*}

Figures~\ref{fig3} and \ref{fig4} display gravitational waves
observed along the rotation axis (i.e., perpendicular to the orbital
plane) for all the models employed in this paper. Figure~\ref{fig3}
is for relatively less massive disks with $15M_\odot \leq M_{\rm disk}
\leq 25M_\odot$ and Fig.~\ref{fig4} is for $30M_\odot \leq M_{\rm
  disk} \leq 50M_\odot$.


For many of the models studied in this paper (the models except for
L12, J13, K14, and M15), the waveforms are characterized by initial
burst waves of a high amplitude and subsequent quasi-periodic waves of
a low amplitude. We refer to this waveform type as type I.  To
classify the waveform, specifically, we first determine the maximum
amplitude of gravitational waves, $h_{\rm max}$, for each model and
define $h_{\rm crit}:=h_{\rm max}/\exp(1)$. Then for each waveform, we
count the wave cycle that satisfies that the amplitude of
gravitational waves is higher than $h_{\rm crit}$. The waveforms with
the wave cycle, $N_{\rm cyc}$, less than 2 are classified as type I. 

The number of the wave cycle of the burst waves is smaller for more
compact or more massive disks for a given value of $c^2r_{\rm
  out}/(GM_{\rm BH,0})$. The maximum amplitude of the burst
gravitational waves is $10^{-22}$--$10^{-21}$ at the hypothetical
distance to the source of $D=100$\,Mpc and is higher for more compact
disks with smaller values of $c^2r_{\rm out}/(GM_{\rm BH})$ or for
more massive disks.  The maximum amplitude is reached when the value
of $\delta_1$ becomes the maximum.  On the other hand, the amplitude
of the quasi-periodic waves emitted in the later stage does not depend
on the value of $c^2r_{\rm out}/(GM_{\rm BH})$ as strongly as that of
the initial burst waves and is always $\alt 10^{-22}$ for
$D=100$\,Mpc.

For the waveforms of models L12, J13, K14 and M15, $N_{\rm cyc} > 2$,
and the amplitude of the initial burst gravitational waves is only by
a factor of 2--3 higher than the subsequent quasi-periodic waves in
contrast to the models of the type I waveform.  We refer to this
waveform type as type II.  For the type II waveform, the maximum
amplitude is $\alt 2\times 10^{-22}$ for $D=100$\,Mpc irrespective of
the disk mass. In terms of $\delta_{1,{\rm max}}$, the type I waveform
is obtained for $\delta_{1,{\rm max}} \agt 0.15$ while the type II
waveform is for $\delta_{1,{\rm max}} \alt 0.15$ (see
Table~\ref{table2}).

We can consider that there are two types among the type I
waveforms. For the models of the very compact or massive disks, the
burst waves are characterized by only one wave of a very high
amplitude (i.e., $N_{\rm cyc} \alt 1$; see the waveforms for models
M12, M13, H14, H15, E15, and E16). We refer to this waveform as type
Is.  The short duration of the highest amplitude phase is due to the
fact that the peak values of $\delta_1$ for these cases are relatively
large and hence by the gravitational torque exerted from the
non-axisymmetric structure, a large amount of the disk matter is
ejected from the high-density region of the disk settling to a less
compact disk in a short timescale.  For other type I models (L11, J12,
K13, M14, H16, E17), the gravitational torque at the moment that
$\delta_1$ is highest is not large enough to expel a large fraction of
matter from the high-density region of the disk. As a result, the
burst waves are emitted for a couple of wave cycles. We refer to the
waveform type as type Iw for this case.  In terms of $\delta_{1,{\rm
    max}}$, the type Iw waveform is obtained only for $\delta_{1,{\rm
    max}} \alt 0.25$. We note that the waveforms for models M13 and
H15 might be classified as type Iw. Thus, we consider that these models
are located near the boarder for distinguishing types Iw and Is (note
that the waveform should change continuously with the change of the
compactness and mass of the disk, and hence, it would not be possible
to definitely classify the waveform into a particular type).

The numerical results presented here illustrate that the compactness
and mass of the disk (which are in reality determined by the angular
momentum profile of the progenitor star prior to the collapse) are the
key parameters for determining the type of the
waveforms. Specifically, for the models employed in this paper, the
type II waveforms are obtained broadly for $M_{\rm disk}/M_\odot < -10
+ c^2r_{\rm out}/(GM_{\rm BH,0})$, and otherwise, the type I waveform
is the result (see Fig.~\ref{fig5}). Also, broadly for $M_{\rm
  disk}/M_\odot > 3 + c^2r_{\rm out}/(GM_{\rm BH,0})$, the waveform
becomes the type Is.  We note that this would be quantitatively valid
only for $c^2r_{\rm out}/(GM_{\rm BH,0}) \geq 20$ and for the initial
angular momentum profile employed in this paper. However, we infer
that qualitatively similar relations would be satisfied irrespective
of the disk configuration for the massive disks.

For the type I waveform, the amplitude of the burst-wave part is
appreciably larger than the quasi-periodic waves subsequently emitted.
Thus, if gravitational waves of this type are detected by
gravitational-wave detectors with a low signal-to-noise ratio, it
would not be very easy to confirm the detection of the quasi-periodic
wave part, and essentially, the attention would be paid only to the
burst-waves part.  For the case of GW190521, the detection with a
sufficiently high signal-to-noise ratio appears to be done only for
the first $\sim 2$--3 wave cycles that have the highest
amplitude~\cite{190521,Nitz}. This waveform is similar to the type Iw
waveform.  This indicates a possibility that the source of GW190521
may not be a merger of high-mass binary black holes but an outcome of
the collapse of a rapidly-rotating very-massive stellar core leading
to a black hole and a massive disk with moderate compactness.  We note
that for a variety of the mass and the compactness of the disk, the
type Iw waveform resembles GW190521; that is, as long as a relation
between the mass and the compactness of the disk is satisfied (see
Fig.~\ref{fig5}), no artificial fine-tuning is necessary to reproduce a
waveform similar to GW190521.

By contrast, for the type II waveform, i.e., for the case of less
compact and less massive disks, the waveforms do not resemble that of
GW190521: For these models, the amplitude of the initial burst waves
is not much higher than the subsequent quasi-periodic waves.  Also,
for the type Is waveform, i.e., for high-mass and compact disk models
such as M12, M13, H14, H15, E15, and E16, the wave cycle of the
high-amplitude burst waves is $\alt 1$, which is not as many as that
of GW190521.


Figure~\ref{fig6} displays the dimensionless effective amplitude of
gravitational waves for the selected models: The left panel compares
the spectra for given values of the disk mass with different
compactness and the right panel compares the spectra for approximately
the same value of $c^2r_{\rm out}/(GM_{\rm BH,0})(\sim 30)$ with
different disk mass.  Here, the effective amplitude is defined from
the Fourier transform of gravitational waves, $h(f)$, as $h_{\rm
  eff}:=|f h(f)|$, where $f$ denotes the frequency of gravitational
waves. The hypothetical distance to the source is 100\,Mpc in this
figure and we assume the observation from the rotation axis (i.e.,
from the most optimistic direction) as in Figs.~\ref{fig3} and
\ref{fig4}.  For the setting with $M_{\rm BH,0}=50M_\odot$,
irrespective of the disk compactness, the frequency at the peak value
of $h_{\rm eff}$ becomes $f_{\rm peak} \approx 40$--50\,Hz which
agrees approximately with that of GW190521. Here the peak frequency is
slightly lower, $\sim 40$\,Hz, for the models with more massive disks
such as H14--H16 and E15--E17.  Our interpretation for this is that
the total mass inside the region for which the disk has the highest
density, $M_{\rm tot}$, is appreciably larger than $M_{\rm BH,0}$, and
thus, the angular velocity of the one-armed spiral mode is slower (in
inversely proportional to $M_{\rm tot}$) for a give compactness of the
disk.

The peak amplitude of $h_{\rm eff}$ is higher for more compact disks
(i.e., for smaller values of $c^2 r_{\rm out}/(GM_{\rm BH,0})$) with a
given value of the disk mass and for more massive disks, as expected
from Figs.~\ref{fig3} and \ref{fig4}.  The noteworthy features found
from Fig.~\ref{fig6} are as follows: Irrespective of the waveform
type, (i) the effective amplitude increases with the frequency for $f
\leq f_{\rm peak}$ and (ii) the effective amplitude steeply drops for
the frequency beyond $f_{\rm peak}$. Although the feature (ii) is
similar to those of binary black-hole mergers, the feature (i) is
totally different from them. Thus, in the detection of gravitational
waves with a high signal-to-noise ratio, it would be easy to
distinguish the waveforms of the massive black hole-disk systems from
those of binary black hole mergers. However, for the detection with a
low signal-to-noise ratio (which is in particular the case for the
low-frequency region of $\alt 30$\,Hz with the gravitational-wave
detectors operated in the third-observational run~\cite{ligo3}), such
difference is unlikely to be confirmed.

\begin{figure*}[tp]
\vspace{-3mm}
\includegraphics[width=80mm]{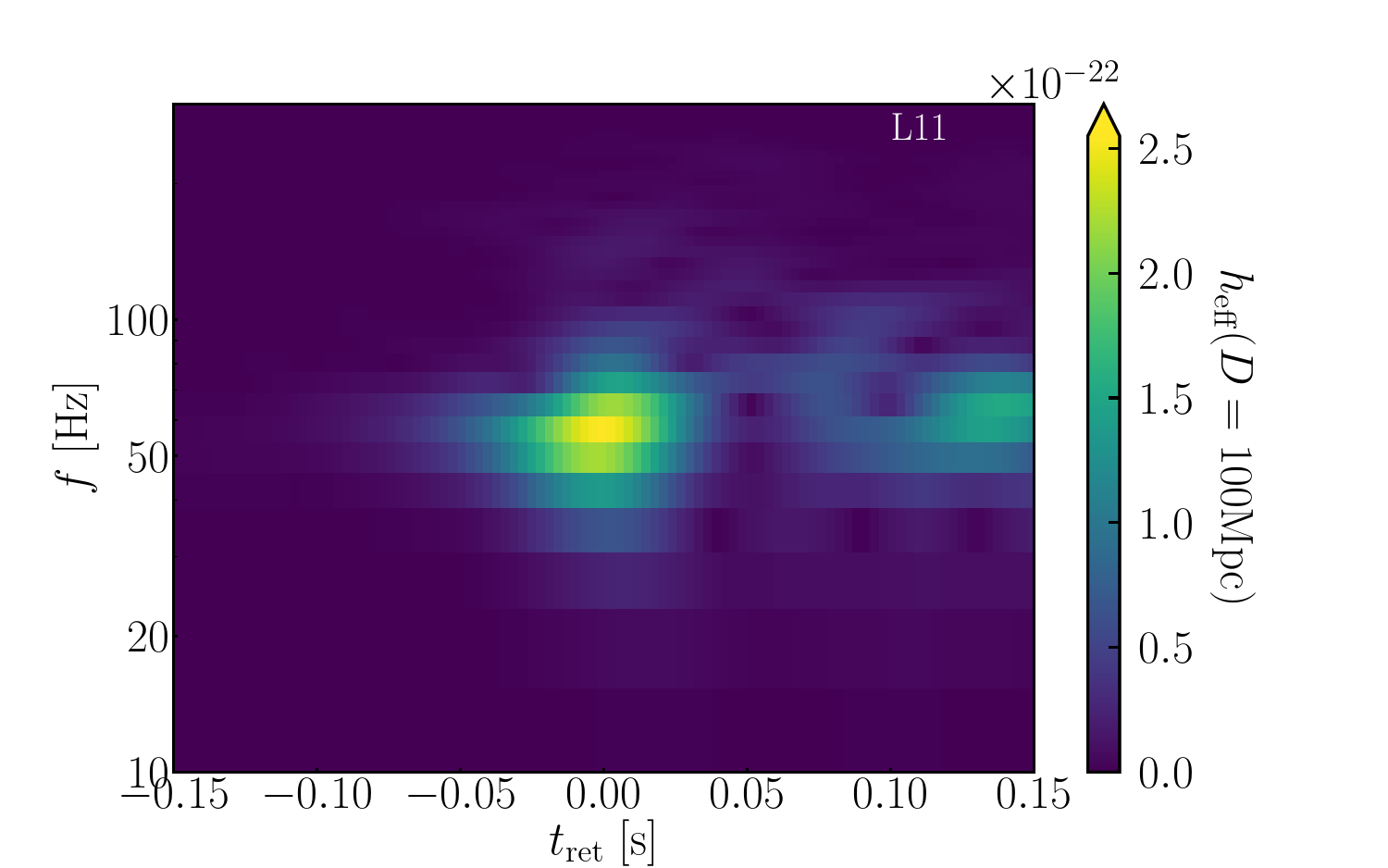}~
\includegraphics[width=80mm]{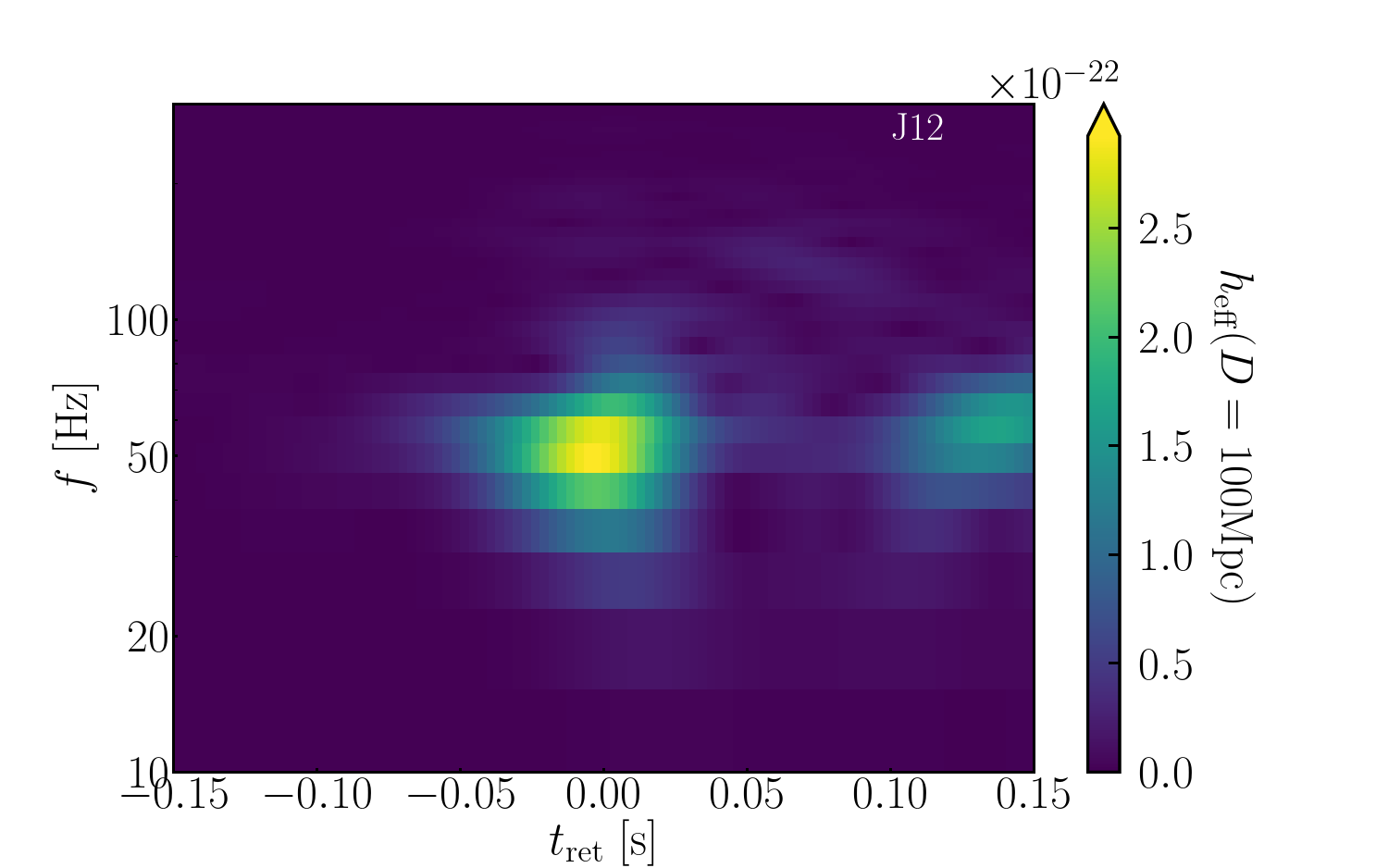} \\
\includegraphics[width=80mm]{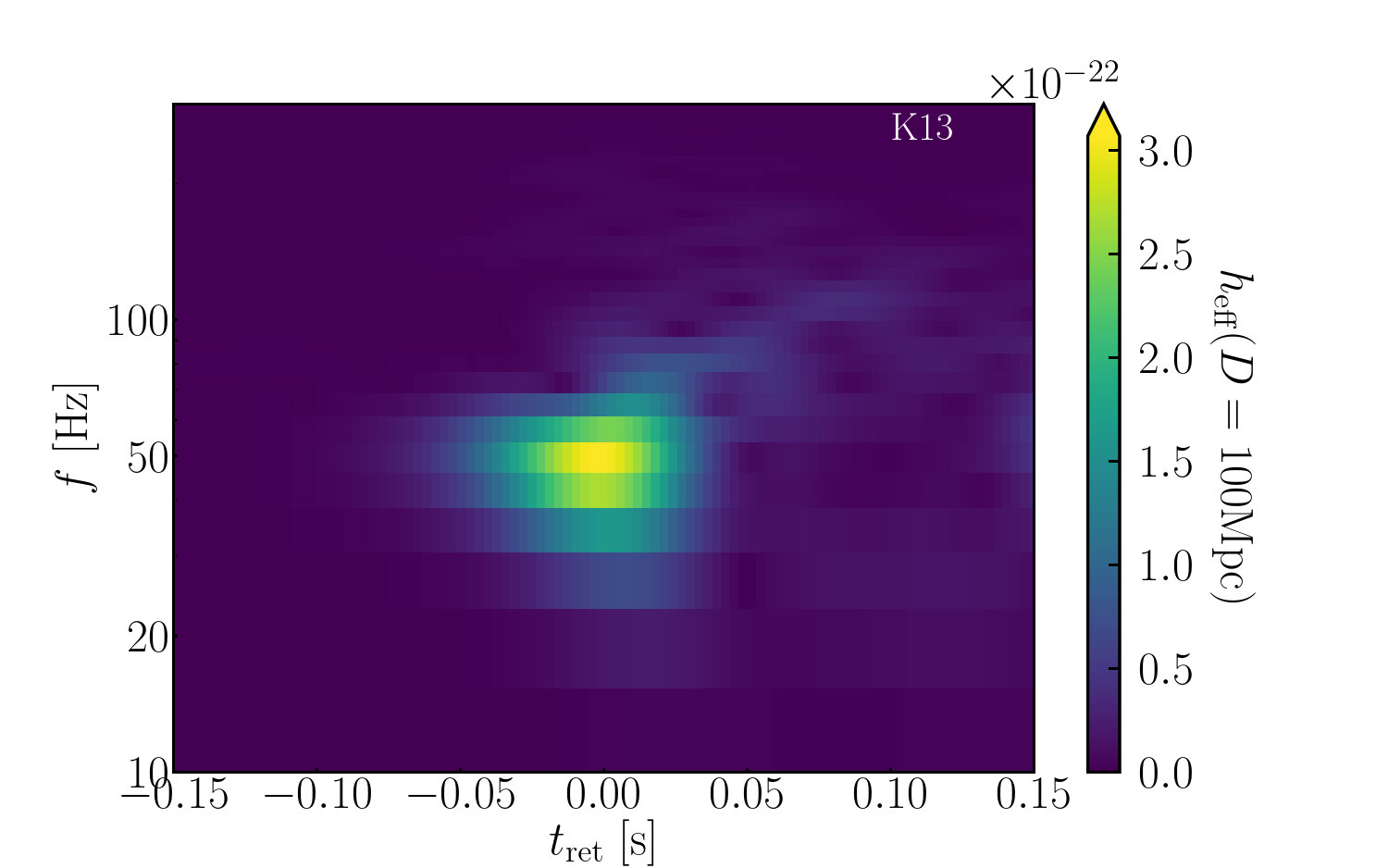}~ 
\includegraphics[width=80mm]{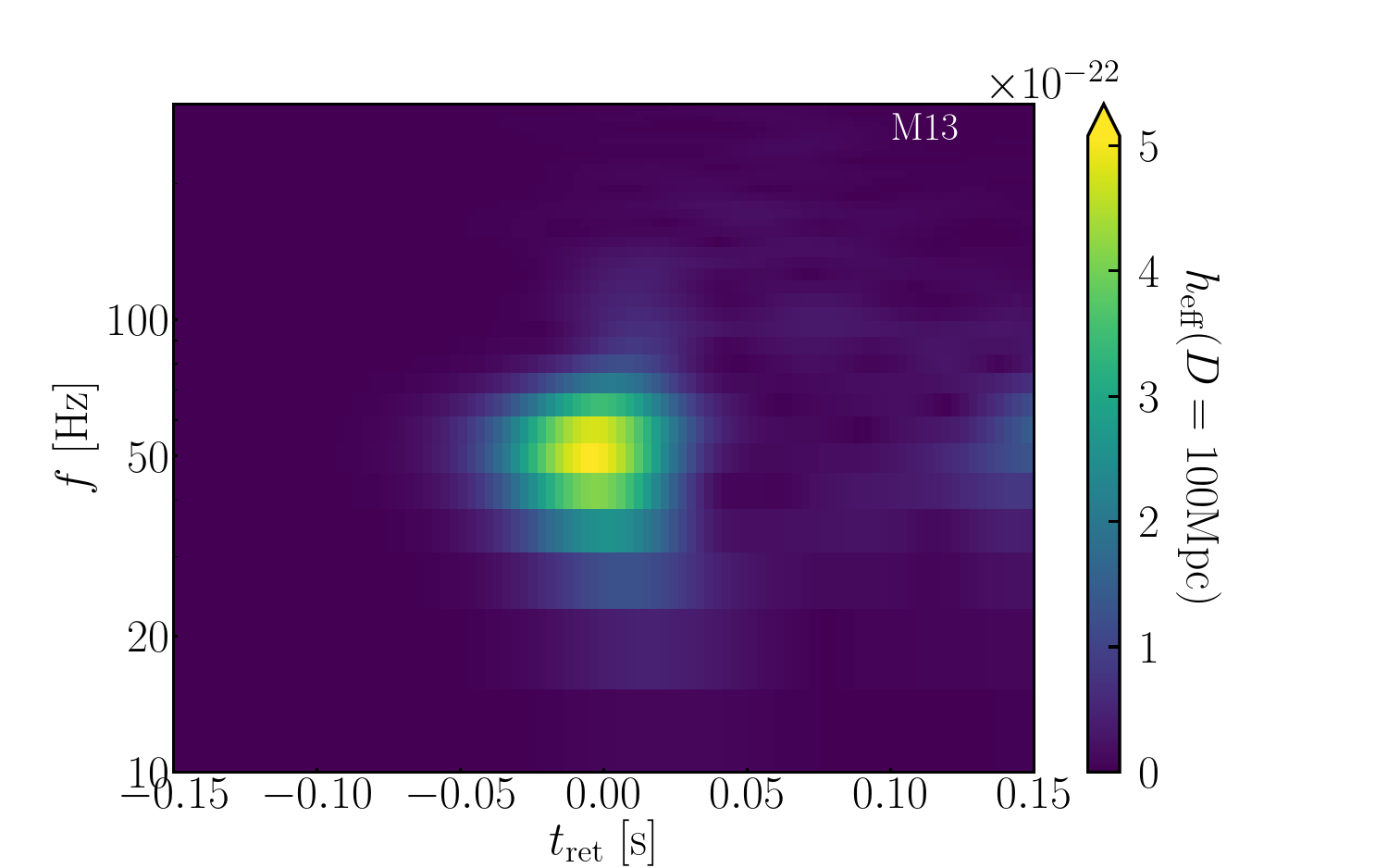} \\
\includegraphics[width=80mm]{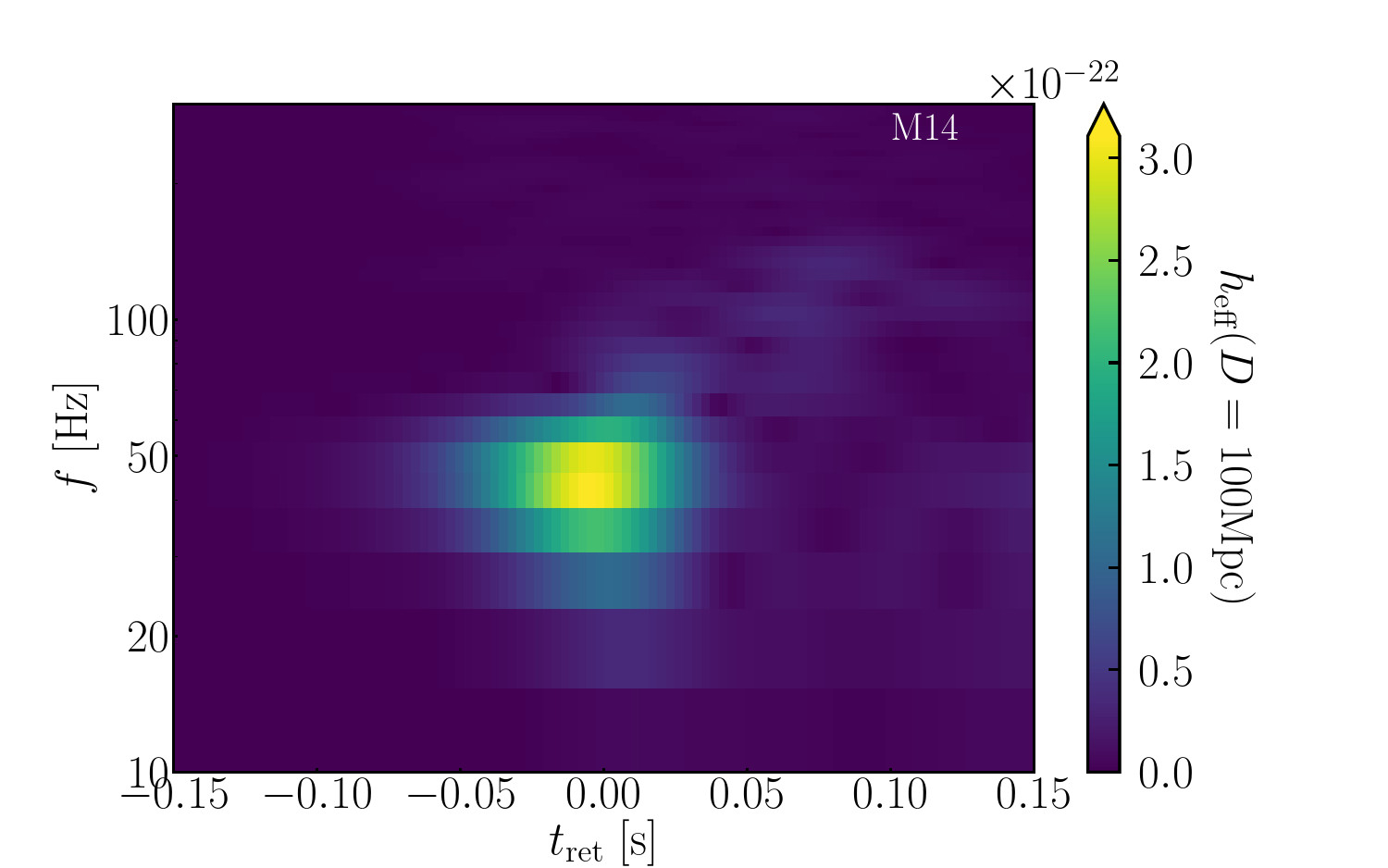}~ 
\includegraphics[width=80mm]{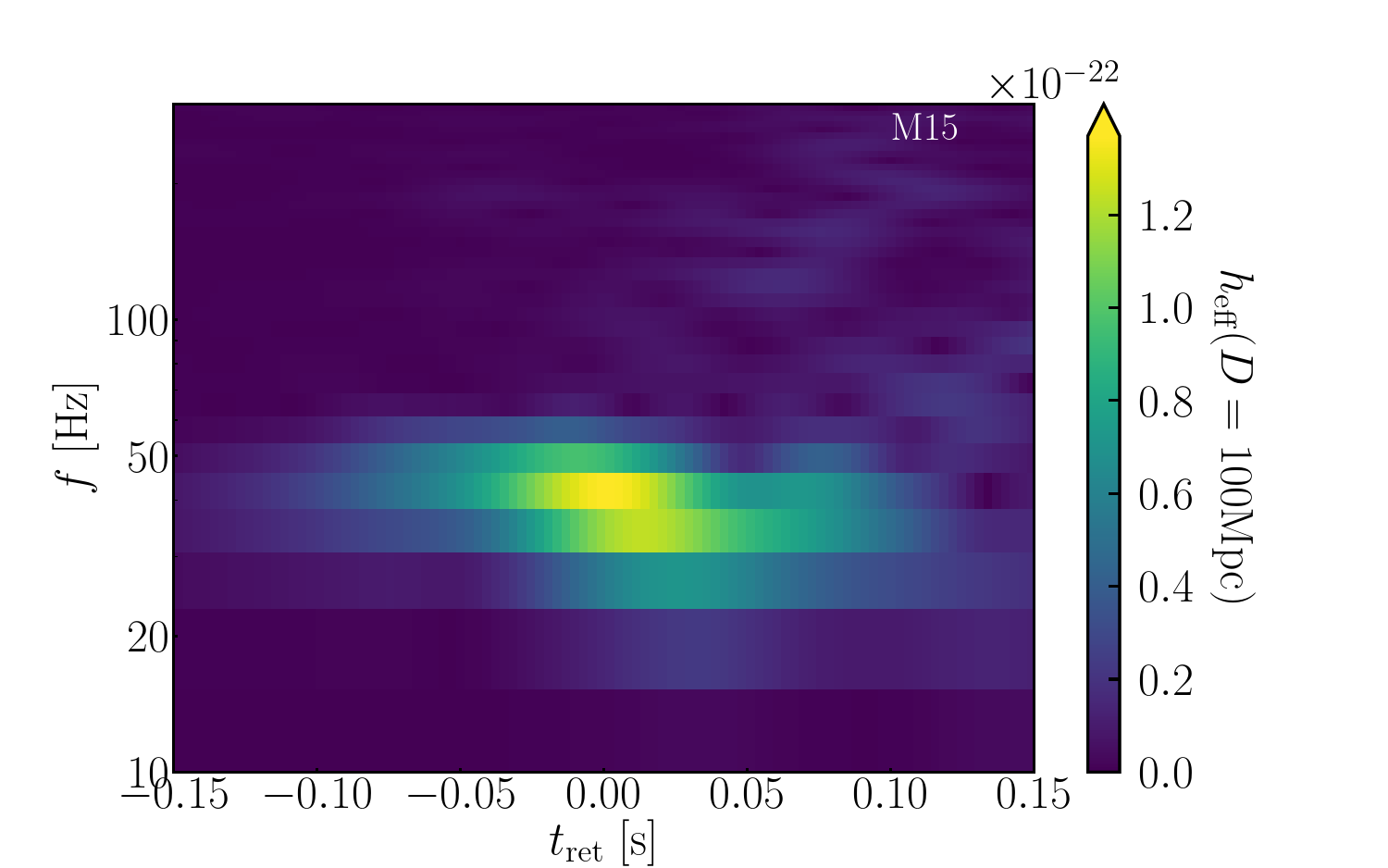} 
\includegraphics[width=80mm]{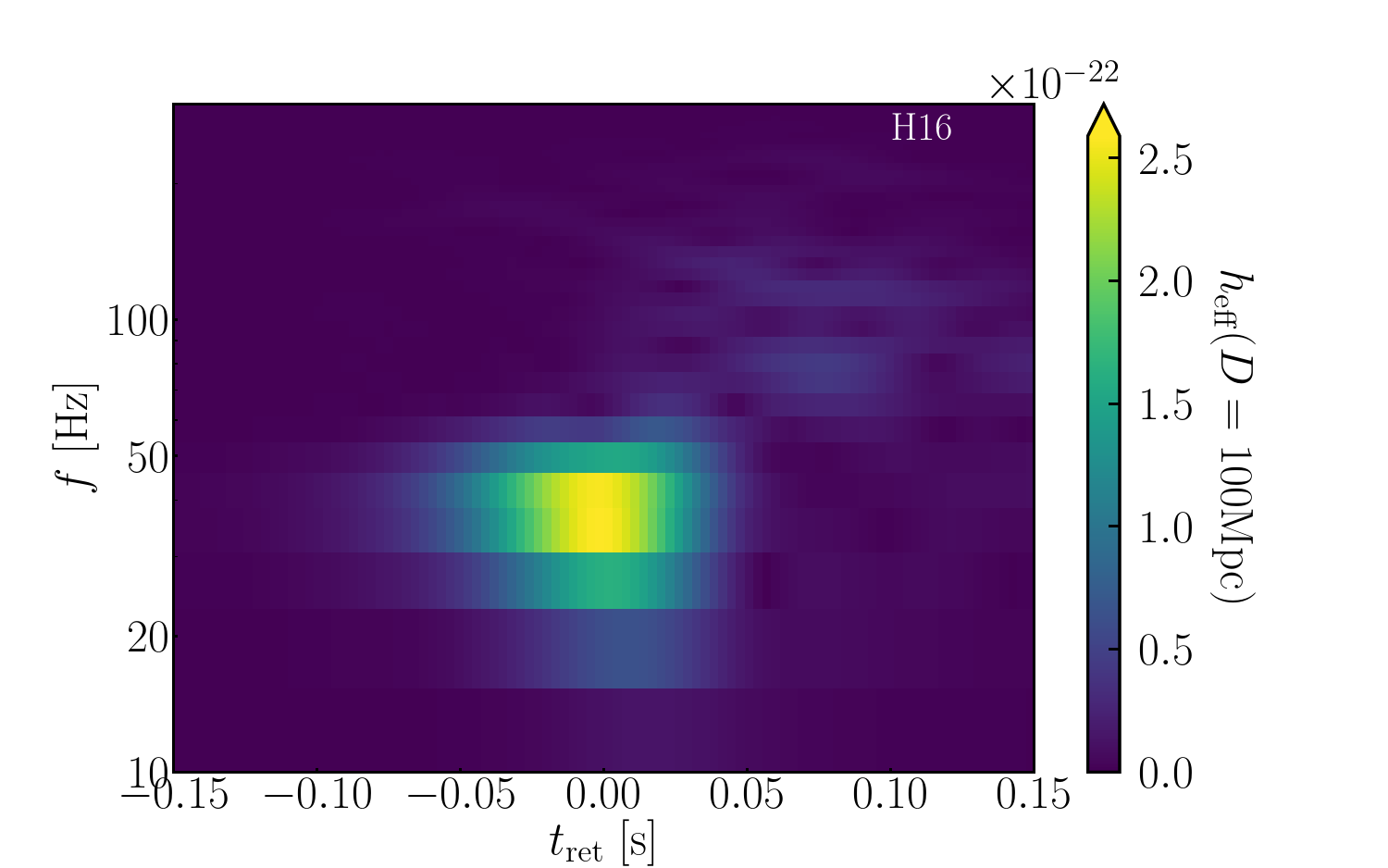}~ 
\includegraphics[width=80mm]{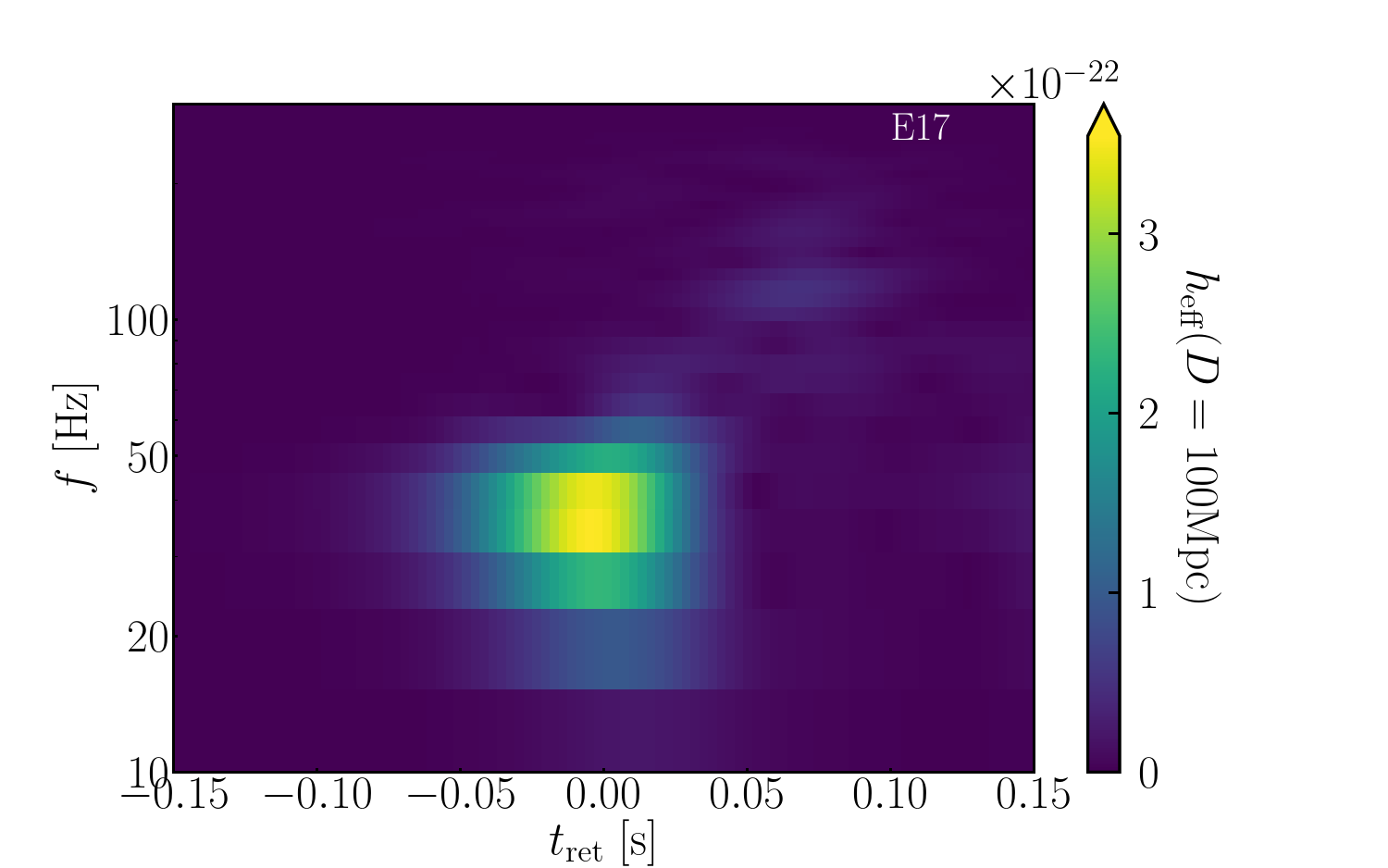} 
\caption{Spectrograms (relation between the dominant frequency as
  a function of time) are displayed for models L11, J12, K13, M13, M14,
  M15, H16, and E17. 
\label{fig7}}
\end{figure*}

Figure~\ref{fig7} plots the spectrogram (dominant frequency as a
function of time) of gravitational waves for models L11, J12, K13,
M13, M14, M15, H16, and E17. As expected from Figs.~\ref{fig3} and
\ref{fig4}, one prominent spot sharply appears at $f \sim
40$--$50$\,Hz for all the models of the type Iw waveform (i.e., except
for M15 of type II waveform). This feature is indeed similar to that
of GW190521 presented in Ref.~\cite{190521} under the condition that
the signal-to-noise ratio is not very high. We note that as shown in
Fig.~\ref{fig6}, for models H16 and E17 for which the disk mass is
comparable to the black-hole mass, the prominent spot appears at a
relatively low frequency with $f \sim 40$\,Hz while for other models,
it is $\sim 50$\,Hz. On the other hand, for model M15 of the type II
waveform, not one prominent spot but an extended bright region appears.
This feature is universally found for models L12, J13, and K14, i.e.,
for the low-mass or less compact models, and is different from that of
GW190521.

The signal-to-noise ratio of gravitational waves with respect to the
designed sensitivity of advanced LIGO at the hypothetical distance of
100\,Mpc is listed in Table~\ref{table2}. Here, the signal-to-noise
ratio is calculated for the detection by a single detector.  Note that
the numerical simulations were stopped at 0.3--0.5\,s after the peak
gravitational-wave amplitude is reached. Since long-lasting
quasi-periodic waves with the duration longer than 0.5\,s is likely to
be present, the signal-to-noise ratio shown here should be considered
as the minimum value.

Taking into account that the sensitivity of advanced LIGO in the third
observational run was by a factor of $\sim 2$ lower than the designed
sensitivity~\cite{ligo3}, the horizon distance to the source for
gravitational waves from the massive black hole-disk systems would be
$\sim 100$--300\,Mpc for advanced LIGO at the third observational
run. Here, for the models with more compact and more massive disks,
the horizon distance is larger.  Thus, if the source of GW190521 was a
massive black hole-disk system that results from a collapse of a
rapidly-rotating very-massive star, it should occur in a fairly nearby
galaxy. Then, we could have a question whether a metal-poor star,
which can form a very massive star, could exist in the relatively
late-time universe.  However, currently, we do not have rich
observational information on this, although this question may be
answered by the future observation~\cite{EMPG}.


It is also worthy to note that in the future detectors currently
proposed, such as Einstein Telescope~\cite{ET} and Cosmic
Explorer~\cite{CE}, the signal-to-noise ratio for the events studied
in this paper will be more than $10$ times as high as the detectors
currently in operation. For such future detectors, the horizon
distance will be deeper than 1\,Gpc and the formation rate of the
very-massive stars is likely to be higher than in the late-time
universe. Considering that the typical frequency of gravitational
waves is approximately proportional to the inverse of the black-hole
mass and that the future detectors will have a high sensitivity down
to $\alt 20$\,Hz~\cite{ET,CE}, the detectability will be also high for
black holes of higher mass of $\sim 100M_\odot$.  Thus, in the future,
gravitational waves not only from binary black hole mergers but also
from high-mass black hole-disk systems (that would result from the
collapse of rapidly-rotating very-massive stars) could be an
interesting target for exploring the formation process and formation
history of massive black holes of mass $\sim 50$--$100M_\odot$.

\section{Discussion}\label{sec4}


Just prior to the formation of the systems composed of a massive black
hole and a massive disk which we considered in this paper, a stellar
core collapse of very-massive star, triggered by the pair instability,
should occur. Since the black hole is likely to be formed immediately
after the collapse~(e.g., Refs.~\cite{uchida1,uchida2}),
electromagnetic counterparts associated with the usual core-collapse
supernovae are unlikely to accompany with this. However, after the
formation of a black hole, the material could be ejected from the
massive disk surrounding the black hole for the long-term evolution,
and such material injects the outgoing momentum and energy into the
extended envelop of radius $\sim 10^{14}$\,cm surrounding the
carbon-oxygen core~\cite{uchida1}. For a sufficiently high energy
injection, the envelop will explode as in the core-collapse
supernovae, and the ejecta may be the source for an electromagnetic
counterpart in the optical-infrared bands with a high luminosity. The
brightness of the supernova-like explosion is likely to depend
strongly on the kinetic energy, $E_{\rm kin}$, of the mass outflow
from the disk~\cite{uchida1}. Our previous work~\cite{Fujiba21} shows
that 10--20\% of the disk material could be ejected by a viscous
process with the typical velocity of $0.05c$ from massive disks
surrounding central black holes.  Assuming these values, we have
\beq
E_{\rm kin} \approx 10^{52}\, {\rm erg} 
\left({M_{\rm eje} \over 5M_\odot}\right)
\left({v_{\rm eje} \over 0.05c}\right)^2,
\eeq
where $M_{\rm eje}$ and $v_{\rm eje}$ are the rest mass and velocity
of the ejecta. Thus for the ejecta mass of 1--$10M_\odot$, the
injection energy is broadly considered to be
$10^{51}$--$10^{52}$\,erg.  We note that the ejecta energy may be
increased by subsequent nucleosynthesis and radio-active decay.

Using the Arnett's model~\cite{Arnett}, Ref.~\cite{uchida1} estimated
that for $E_{\rm kin}=10^{51}$--$10^{52}$\,erg and the envelop mass of
$\sim 150M_\odot$, the absolute luminosity of an explosion would be
$\sim 10^{42}$--$10^{43}$\,erg/s with the duration of $O(1\,{\rm
  yr})$.  Thus, for a hypothetical distance of 100\,Mpc, the apparent
magnitude is $\sim 17$--19\,mag. Such a source is observable or
excludable by the current optical-infrared telescopes for the
transient electromagnetic counterpart search as indicated in
Ref.~\cite{EM190521}.  Unfortunately, the electromagnetic counterparts
of GW190521 were not seriously searched by the optical-infrared
telescopes, because GW190521 was announced as a candidate for binary
black holes.


The collapse of rapidly-rotating very-massive stars of initial mass
larger than $\sim 200$--$300M_\odot$ would be rare and hence the typical
distance to this collapse event would be large $\gg 10$\,Mpc. Thus,
the possible electromagnetic counterparts of this kind of the event
are not likely to be very luminous, and it might not be easy to detect
it in the absence of the information of the sky localization. As we
studied in this paper, gravitational waveforms from the collapse of
rapidly-rotating very-massive stars could be similar to that of the
merging binary black holes with a value of high chirp mass $\agt
60M_\odot$. Thus an alert for such an event could be issued as a
candidate for a binary black hole merger as in the case of GW190521 in
the future. Although a careful follow-up observation by
electromagnetic telescopes is usually absent for the candidates
announced as binary black holes, in the future, it will be interesting
to perform follow-up observations for the candidates of very high-mass
binary black holes. The problem is that quite unfortunately the
LIGO-Virgo collaboration never announce the chirp mass (i.e., typical
frequency) of the candidate events of gravitational waves. Thus it
will not be possible to distinguish the interesting events among a
huge amount of the ``binary black hole'' candidates unless the policy of
the LIGO-Virgo collaboration is changed in the future.

In this paper, we focus only on the excitation of an unstable mode in
massive disks surrounding a high-mass black hole and associated
gravitational waves emitted.  Another interesting aspect of this
system is that a large amount of matter could be ejected from the
massive disk during the enhancement of the one-armed spiral density
wave and probably through the subsequent long-term viscous
evolution. For a reliable estimation of $E_{\rm kin}$ mentioned above,
it is important to perform a long-term simulation for exploring the
mass ejection process~\cite{Fujiba20,Fujiba21} and to quantitatively
explore the value of $E_{\rm kin}$.  In the ejecta from the disk,
nucleosynthesis could also occur and the radio-active energy could be
released for the additional source of the electromagnetic
counterparts. If an amount of Ni with mass larger than $\sim 1M_\odot$
is synthesized in the ejecta, the event may be similar to the luminous
supernovae (e.g., Ref.~\cite{Maeda2003}). The detailed exploration of
the mass ejection and nucleosynthesis is an interesting topic for the
subsequent work. It is also interesting to explore how much mass can
be ejected from the system. If a significant fraction of the matter
forms a massive disk and is subsequently ejected from the system, the
final mass of the formed black hole can be smaller than the progenitor
mass, and may be smaller than $120M_\odot$, i.e., may come into the
mass-gap range.

There is another possible phenomenon for emitting gravitational waves
similar to GW190521. In this paper, we suppose that a black hole is
formed during the collapse. If the centrifugal force in the central
region of the collapsing very-massive stars is strong enough, not a
black hole but a very massive spheroidal or toroidal object can be
formed (see, e.g., Refs.~\cite{SS05,Zink07,F2021} for other contexts).  If
such an outcome is very compact and dynamically unstable to one-armed
spiral mode or bar-mode deformation, gravitational waves with a high
amplitude, which would have the same order of magnitude as that from
massive black hole-disk systems, can be emitted. Assuming the
formation of a compact object, the estimated frequency of
gravitational waves is
\beqn
f&=& {1 \over \pi}\sqrt{{GM_{\rm obj} \over R_{\rm obj}^3}}
  \nonumber \\
  &\approx & 58\,{\rm Hz} \left({M_{\rm obj} \over 100M_\odot}\right)^{-1}
  \left({R_{\rm obj}\over 5GM_{\rm obj}c^{-2}}\right)^{-3/2},
\eeqn
where $M_{\rm obj}$ is the mass of the spheroidal/toroidal object and
$R_{\rm obj}$ is the typical radius for the highest-density part of
such an object at the onset of the dynamical instability,
respectively.  Thus, the formation of a compact spheroid/toroid from a
rapidly rotating very-massive stellar core can be another scenario for
GW190521. 

Finally, we note that for forming a rapidly-rotating very-massive
star, a metal-poor star would be necessary. Such a star is unlikely to
be formed frequently in the late-time universe, and thus, it would be
rare that the collapse of the rapidly-rotating very-massive star is
observed at the distance of several hundred\, Mpc.  However, for given
uncertainty on the formation rate of such stars in metal-poor
galaxies~\cite{EMPG}, it is important to impose the constraint using
the observational results. From this perspective, the
gravitational-wave observation together with the
electromagnetic-counterpart search can provide us a valuable
opportunity.

To summarize, in this paper, we have suggested that burst
gravitational waves of frequency $\sim 50$\,Hz, which are observed
with a small signal-to-noise ratio and are composed only of a few
cycles of high-amplitude waves, can be interpreted not only by those
of binary black holes but by those by other scenarios. We suspect that
the source of GW190521 might not be a merger of binary black holes but
a stellar collapse of a very massive star leading temporarily to a
black hole of mass $\sim 50M_\odot$ and a massive disk of several tens
of solar mass that is dynamically unstable to the one-armed
spiral-shape deformation. The point of this scenario is that the
progenitor should be heavy enough to form a high-mass black hole of
mass $\sim 50M_\odot$ and a high-mass disk, which can emit
gravitational waves of low frequency $\sim 50$\,Hz, although we do not
need the fine-tuning of the disk mass. One should keep in mind that it
is not easy to definitely conclude the source of this-type of burst
events only from a gravitational-wave observation with a low
signal-to-noise ratio.  In the future, electromagnetic counterpart
searches will play a key role for pining down the most probable
scenario and an alert from gravitational-wave observation community
suitable for such searches is obviously necessary.

In the present paper, we prepare an initial condition of a massive
black hole and a massive disk supposing that it would be the outcome
formed during the collapse of a rapidly-rotating very-massive stellar
core. Obviously, this setting is idealized. For more realistic work,
it is necessary to perform a simulation started from a
rapidly-rotating very-massive progenitor star. We plan to perform such
simulations in the subsequent work.


\acknowledgments

We thank Koh Takahashi, Masaomi Tanaka, and Takami Kuroda for helpful
conversations. This work was in part supported by Grant-in-Aid for
Scientific Research (Grant Nos.~JP16H02183, JP18H01213, and
JP20H00158) of Japanese MEXT/JSPS.  Numerical computations were
performed on Sakura and Cobra clusters at Max Planck Computing and
Data Facility.

\appendix
\section{Profiles of the initial models}

\begin{figure*}[t]
\includegraphics[width=55mm]{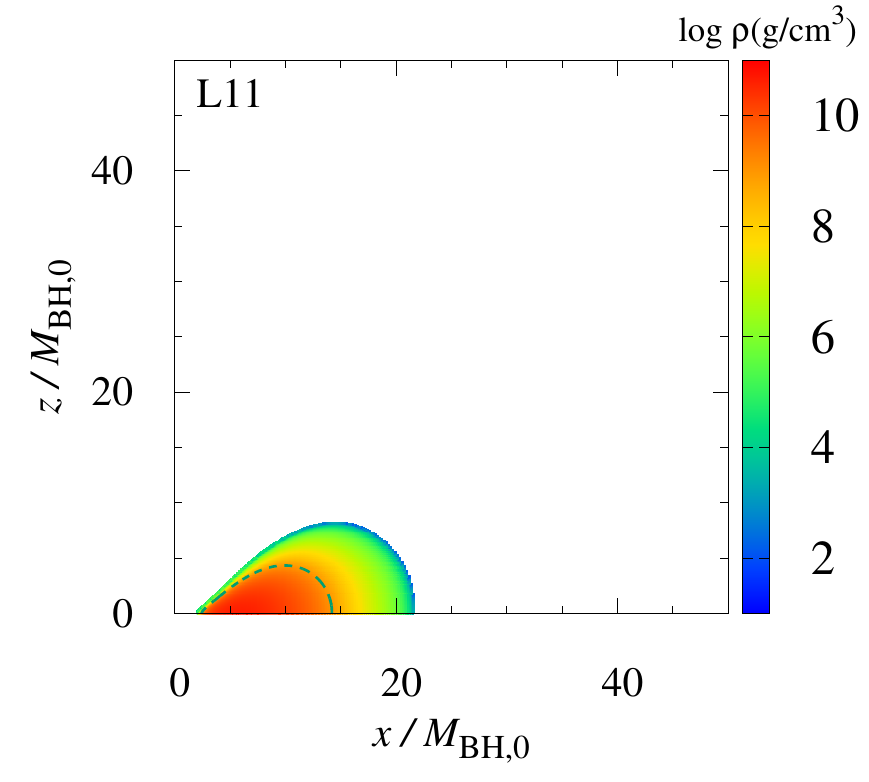}~~
\includegraphics[width=55mm]{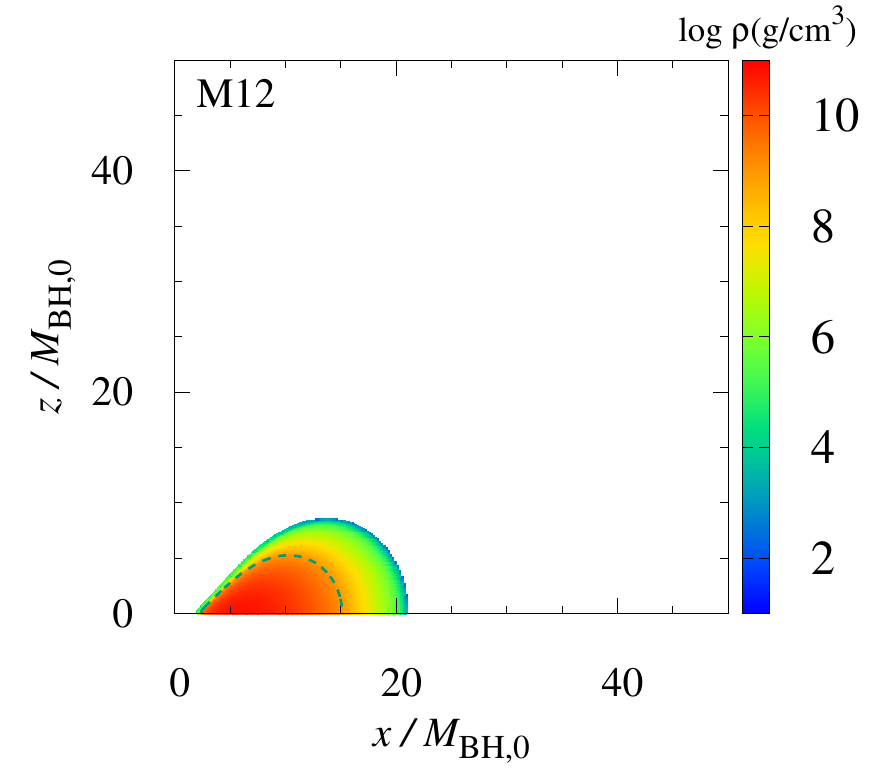}~~
\includegraphics[width=55mm]{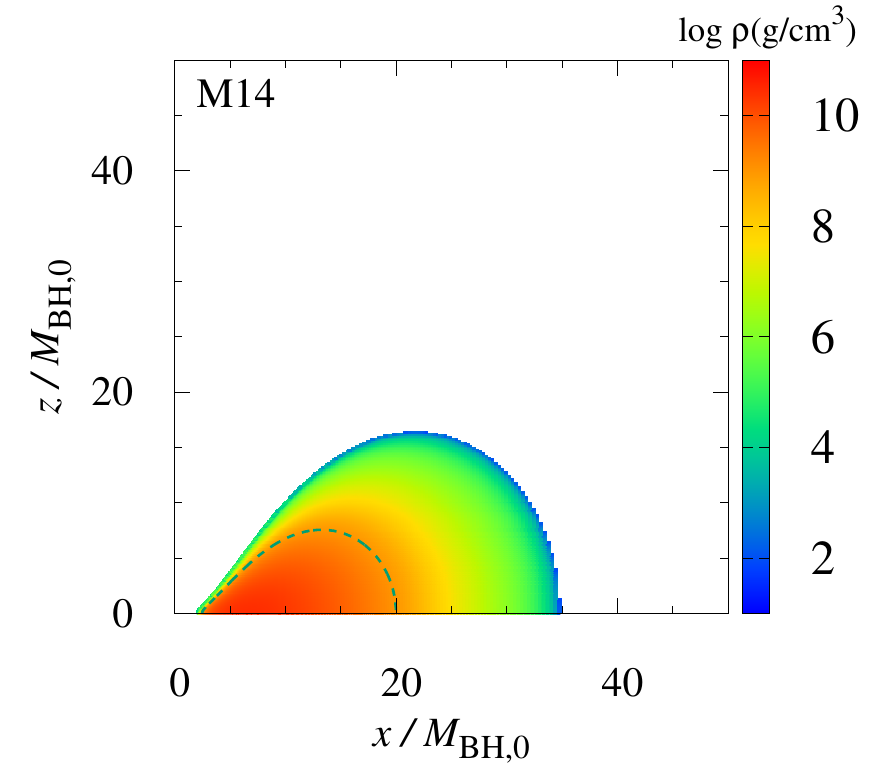}~~
\includegraphics[width=55mm]{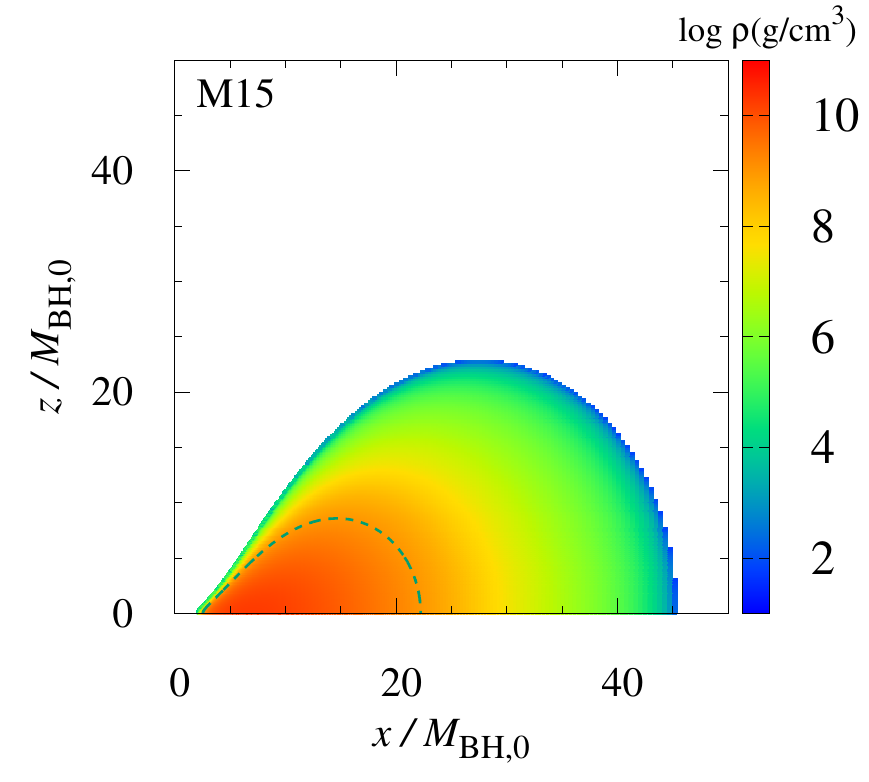}~~
\includegraphics[width=55mm]{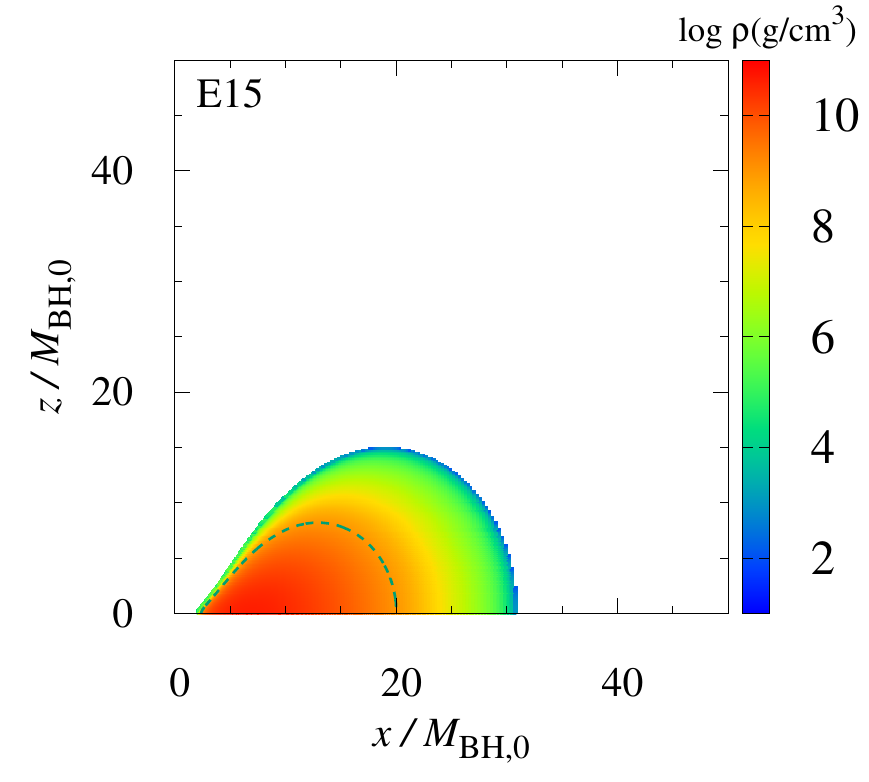}~~
\includegraphics[width=55mm]{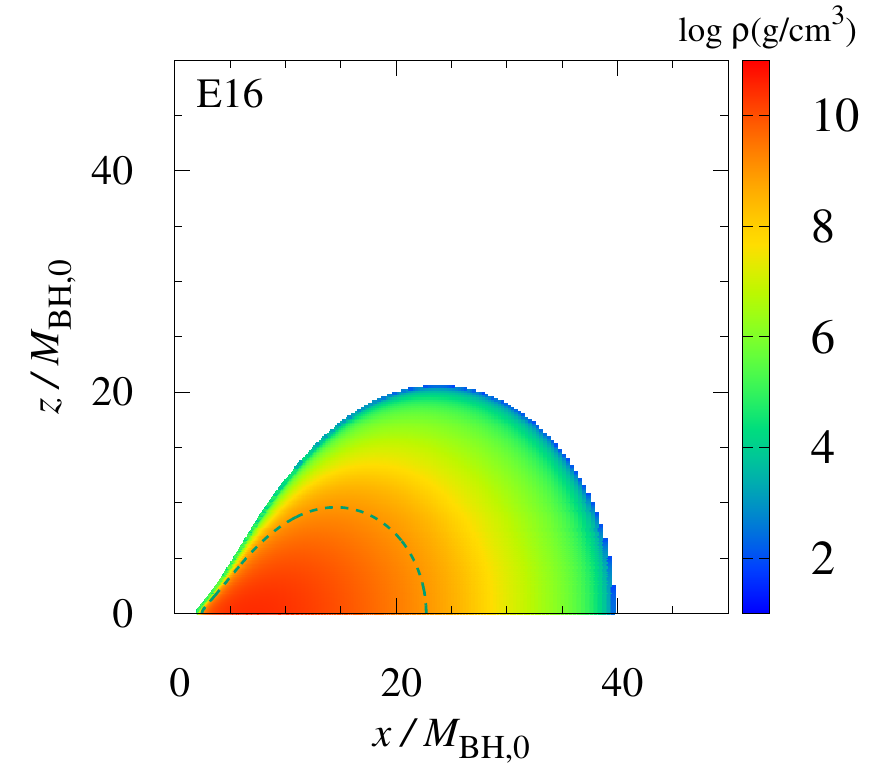}~~
\caption{The density profile for models L11, M12, M14, M15, E15, and
  E16.  The horizontal and vertical axes show $c^2 x/(GM_{\rm BH,0})$
  and $c^2 z/(GM_{\rm BH,0})$, respectively (i.e., we omit $c$ and $G$
  in the labels). The dashed curves in each panel show the location of
  $\rho=10^9\,{\rm g/cm^3}$. 
\label{figA1}}
\end{figure*}

\begin{figure}[t]
\includegraphics[width=84mm]{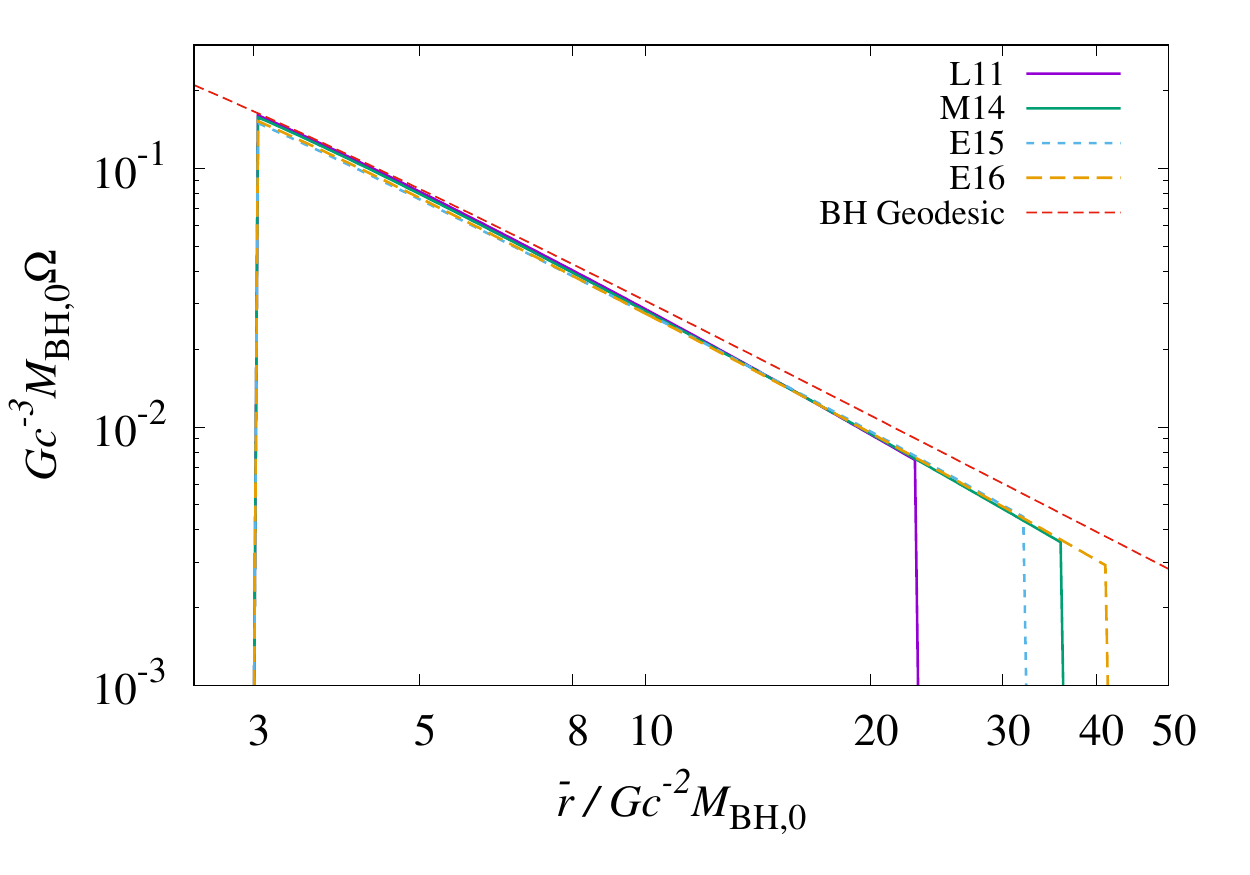}
\caption{The angular-velocity profile for models L11, M14, E15, and
  E16 together with that for the circular orbits around a Kerr black
  hole of mass $M_{\rm BH,0}$ and of dimensionless spin $\chi=0.8$.
  The horizontal axis is $c^2 \bar r/(GM_{\rm BH,0})$ (for the
  definition of $\bar r$ see the text). In this coordinate, the inner
  edge of the disk is located at $c^2 \bar r/(GM_{\rm BH,0})\approx
  3$.
\label{figA2}}
\end{figure}

In Figs.~\ref{figA1} and \ref{figA2}, we display the profiles of the
rest-mass density and angular velocity on the equatorial plane,
respectively, for selected models. Figure~\ref{figA1} shows that the
maximum vertical height of the disk is about half of the extent of the
disk ($c^2 r_{\rm out}/(GM_{\rm BH,0})$) irrespective of the disk mass
and the value of $r_{\rm out}$. Paying attention to the high-density
part of the disks with $\rho \geq 10^9\,{\rm g/cm^3}$, the maximum
ratio of the scale height to the cylindrical radius, $H/R$, is
approximately between $1/2$ and $2/3$. This ratio is smaller for less
massive disk with the same value of $r_{\rm out}/M_{\rm BH,0}$ (for
example, compare models L11 and M12 or M15 and E16), although the
difference is not very appreciable. Comparing models L11 and M12 of
approximately the same value of $c^2r_{\rm out}/(GM_{\rm BH,0})$, we
may say that the disk mass is determined primarily by the value of
$s/k$ since the shape and extent of these disks are similar to each
other.  For higher disk-mass models, the maximum density is located at
a larger value of $c^2x/(GM_{\rm BH,0})$, reflecting the stronger
effect of the self-gravity of the disk.

Figure~\ref{figA2} displays the angular velocity on the equatorial
plane in units of $GM_{\rm BH,0}/c^3$ for models L11, M14, E15, and
E16. Note that for other models the profile is similar to each
other. In this plot, we choose a Boyer-Lindquist radial coordinate
$\bar r$ for the horizontal axis, which is defined from the
quasi-isotropic coordinate, $r$, employed in our study by
\beqn
\bar r=r\left(1+{GM_{\rm BH,0} (1-\chi) \over 2c^2 r} \right)
\left(1+{GM_{\rm BH,0} (1+\chi)\over 2c^2 r} \right),
\nonumber \\
\eeqn
where the dimensionless spin $\chi=0.8$ in the present paper. In
addition to the numerical results for the angular velocity we also
plot the angular velocity for circular orbits of a point mass around
the Kerr black hole, i.e., $Gc^{-3}M_{\rm BH,0}\Omega=[(c^2 \bar
  r/(GM_{\rm BH,0}))^{3/2}+\chi]^{-1}$~\cite{ST83}.

Figure~\ref{figA2} shows that the angular velocity profile depends
only weakly on the mass and the extent of the disk ($c^2 r_{\rm
  out}/(GM_{\rm BH,0})$).  Irrespective of the mass and the extent of
the disks, the angular velocity profile agrees approximately with that
of the circular orbits around the Kerr black hole near the innermost
region including the location of the highest rest-mass density.  This
is reasonable because the orbital motion near the black hole is
determined primarily by the gravity of the black hole.  For the outer
region, the angular velocity ($Gc^{-3} M_{\rm BH,0}\Omega$) is smaller
than that of the circular orbits around the Kerr black hole. A part of
the reason for this is that the pressure force contributes
significantly to supporting the gravity by the black hole and
disk. The other reason is in our choice of the relation between $j$
and $\Omega$: for smaller values of $n$, the angular velocity
decreases more steeply with the increase of the orbital radius. We
confirmed this point by comparing the profile of the angular velocity
for equilibria with $n=1/5$, $1/6$, $1/7$, $1/10$, and $1/20$. This
effect is as significant as that by the pressure force of the disk.


\begin{thebibliography}{99}

\bibitem{190521} R. Abbott et al., Phys. Rev. Lett. {\bf 125}, 101102 (2020).

\bibitem{woosley17} S. E. Woosley, Astrophys. J. {\bf 836}, 244 (2017).

\bibitem{woosley19} S. E. Woosley, Astrophys. J. {\bf 878}, 49 (2019).

\bibitem{Yoshida16} T. Yoshida, H. Umeda, K. Maeda, and T. Ishii, Mon. Not. R. Astron. Soc.
{\bf 457}, 351 (2016). 

\bibitem{190521a} G. Fragione, A. Loeb, and F. A. Rasio,
  Astrophys. J. {\bf 902}, L26 (2020).

\bibitem{190521b} E. J. Farrell et al., arXiv:2009.06585.

\bibitem{190521c} 
T. Kinugawa, T. Nakamura, and H. Nakano, Mon. Not. R. astron.
Soc, {\bf 501}, L49 (2021).

\bibitem{190521d} 
M. Safarzadeh and Z. Haiman, Astrophys. J.
{\bf 903}, L21 (2020).

\bibitem{190521e} B. Lie and V. Bromm, Astrophys. J. {\bf 903},
  L40 (2020). 

\bibitem{Nitz} A. H. Nitz and C. D. Capano, Astrophys. J. {\bf 907}, L9 (2021). 

\bibitem{SS05} M. Shibata and Y. Sekiguchi, Phys. Rev. D {\bf 71}, 024014 (2005). 

\bibitem{UN02} H. Umeda and K. Nomoto, Astrophys. J. {\bf 565}, 385 (2002). 

\bibitem{HW02} A. Heger and S. E. Woosley, Astrophys. J. {\bf 567}, 532 (2002). 

\bibitem{Takahashi16} K. Takahashi, T. Yoshida, H. Umeda, K. Sumiyoshi, and S. Yamada, Mon. Not.
  R. Astron. Soc. {\bf 456}, 1320 (2016).
  
\bibitem{Takahashi18} K. Takahashi, T. Yoshida, and H. Umeda, 
Astrophys. J. {\bf 857}, 111 (2018).

\bibitem{FWH01} C. L. Fryer, S. E. Woosley, and A. Heger, Astrophys. J. {\bf 550}, 372
  (2001). 

\bibitem{Yoon2012} S.-C. Yoon, A. Dierks, and N. Langer, Astron. Astrophys.
  {\bf 542}, A113 (2012). 

\bibitem{Yoon2015} S.-C. Yoon, J. Kang, and A. Kozyreva,
  Astrophys. J. {\bf 802}, 16 (2015). 
  
\bibitem{uchida1} H. Uchida, M. Shibata, K. Takahashi, and T. Yoshida, Astrophys. J.
  {\bf 870}, 98 (2019). 

\bibitem{uchida2} H. Uchida, M. Shibata, K. Takahashi, and T. Yoshida,
  Phys. Rev. D {\bf 99}, 041302 (2019).


 
\bibitem{Hawley91} J. F. Hawley, Astrophys. J. {\bf 381}, 496 (1991). 

\bibitem{Zink07} B. Zink, N. Stergioulas, I. Hawke, C. D. Ott, E. Schnetter, and E. M{\"u}ller, 
Phys. Rev. D {\bf 76}, 024019 (2007). 

\bibitem{Oleg11} O. Korobkin, E. B. Abdikamalov, E. Schnetter,
  N. Stergioulas, and B. Zink, Phys. Rev. D {\bf 83}, 043007 (2011).

\bibitem{Kiuchi11} K. Kiuchi, M. Shibata, P. J. Montero, and
  J. A. Font, Phys. Rev. Lett. {\bf 106}, 251102 (2011).

\bibitem{Wessel} E. Wessel, V. Paschalidis, A. Taokaros, M. Ruiz,
  and S. L. Shapiro, arXiv: 2011.04077. 
 
\bibitem{Bugli} M. Bugli, J. Guilet, E. M{\"u}ller, L. Del Zanna,
  N. Bucciantini, and P. J. Montero, Mon. Not. R. Astron. Soc. {\bf 475}, 108
  (2018). 

\bibitem{Ott11} C. Reisswig, C. D. Ott, E. Abdikamalov, R. Haas,
  P. M\"oesta, and E. Schnetter, Phys. Rev. Lett. {\bf 111}, 151101
  (2013).
  
\bibitem{Fujiba20} S. Fujibayashi, M. Shibata, S. Wanajo, K. Kiuchi, K. Kyutoku, 
and Y. Sekiguchi, Phys. Rev. D {\bf 101}, 083029 (2020).

\bibitem{Fujiba21} S. Fujibayashi, M. Shibata, S. Wanajo, K. Kiuchi, K. Kyutoku, 
and Y. Sekiguchi, Phys. Rev. D {\bf 102}, 123014 (2020). 

\bibitem{BSSN} M. Shibata and T. Nakamura, Phys. Rev. D {\bf 52},
  5428(1995): T. W. Baumgarte and S. L. Shapiro, Phys. Rev. D {\bf
    59}, 024007(1998). 

\bibitem{puncture} M. Campanelli, C. O. Lousto, P. Marronetti, and
  Y. Zlochower, Phys. Rev. Lett. {\bf 96}, 111101 (2006): J. G. Baker,
  J. Centrella, D.-I. Choi, M. Koppitz, and J. van Meter,
  Phys. Rev. Lett. {\bf 96}, 111102 (2006).

\bibitem{Z4c} D. Hilditch, S. Bernuzzi, M. Thierfelder, Z. Cao, W. Tichy, 
and B. Br{\"u}gmann, Phys. Rev. D {\bf 88}, 084057 (2013). 

\bibitem{Fujiba2018} S. Fujibayashi, K. Kiuchi, N. Nishimura, Y. Sekiguchi, 
and M. Shibata, Astrophys. J {\bf 860}, 64 (2018). 

\bibitem{Fujiba2020} S. Fujibayashi, S. Wanajo, K. Kiuchi, K. Kyutoku,
  Y. Sekiguchi, and M. Shibata, Astrophys. J. {\bf 901}, 122 (2020). 

\bibitem{K10} M. Shibata and K. Taniguchi, Phys. Rev. D {\bf 77},
  084015 (2008); K. Kyutoku, M. Shibata, and K. Taniguchi,
  Phys. Rev. D {\bf 82}, 044049 (2010).
  
\bibitem{SACRA} T. Yamamoto, M. Shibata, and K. Taniguchi, Phys. Rev. D
  {\bf 78}, 064054 (2008).

\bibitem{RP11} C. Reisswig and D. Pollney,
  Class. Quant. Grav. 28, 195015 (2011). 

\bibitem{DD2} S. Banik, M. Hempel, and D. Bandyophadyay,
        Astrophys. J. Suppl. Ser. {\bf 214}, 22 (2014).

\bibitem{Timmes} F. X. Timmes and F. D. Swesty,
  Astrophys. J. Suppl. {\bf 126}, 501 (2000).

\bibitem{Shibata2007} M. Shibata, Phys. Rev. D {\bf 76}, 064035 (2007). 


\bibitem{Sekig15} Y. Sekiguchi, K. Kiuchi, K. Kyutoku, and M. Shibata,
Phys. Rev. D {\bf 91}, 064059 (2015). 

\bibitem{Sekig16} Y. Sekiguchi, K. Kiuchi, K. Kyutoku, M. Shibata, 
and K. Taniguchi, Phys. Rev. D {\bf 93}, 124046 (2016). 

\bibitem{Kyutoku18} K. Kyutoku, K. Kiuchi, Y. Sekiguchi, M. Shibata, 
and K. Taniguchi, Phys. Rev. D {\bf 97}, 023009 (2018). 

\bibitem{BH98} S. A. Balbus and J. F. Hawley, Rev. Mod. Phys. {\bf 70}, 
  1 (1998).

\bibitem{Hawley11} J. F. Hawley, X. Guan, and J. H. Krolik,
  Astrophys. {\bf 738}, 84 (2011).

\bibitem{SS73} N. I. Shakura and R. A. Sunyaev, Astron. Astrophys.
  {\bf 24}, 337 (1973). 

\bibitem{ligo3} R. Abbott et el., arXiv:2010.14527. 
  
\bibitem{EMPG} T. Kojima et al. arXiv: 2006.03831. 

\bibitem{ET} M. Punturo et al., Class. Quantum. Grav. {\bf 27}, 194002 (2010).

\bibitem{CE} B. P. Abbott et al. Class. Quantum. Grav. {\bf 34}, 044001 (2017). 
  

\bibitem{Arnett} W. D. Arnett, Astrophys. J. {\bf 237}, 541 (1980).

\bibitem{EM190521} M. J. Graham et al., Phys. Rev. Lett. {\bf 124}, 251102 (2020). 

\bibitem{Maeda2003} K. Maeda and K. Nomoto, Astrophys. J. {\bf 598},
  1163 (2003).

\bibitem{F2021} S. Fujibayashi, K. Takahashi, Y. Sekiguchi, and M. Shibata,
  Astrophys. J. in submission (arXiv: 2102.04467).
  
\bibitem{ST83} E.g., S. L. Shapiro and S. A. Teukolsky,
{\em Black Holes, White Dwarfs, and Neutron Stars}, Wiley Interscience
(New York, 1983).

\end{thebibliography}
\end{document}